\documentstyle[12pt,amssymb,epsfig]{article}  %Latex 2.09
\input scrload
\newcommand{\newsection}[1]{
\addtocounter{section}{1} \setcounter{equation}{0}
\setcounter{subsection}{0} \addcontentsline{toc}{section}{\protect
\numberline{\arabic{section}}{{\rm #1}}} \vglue .6cm \pagebreak[3]
\noindent{\bf  \thesection. #1}\nopagebreak[4]\par\vskip .3cm}
\newcommand{\newsubsection}[1]{
\addtocounter{subsection}{1}
\addcontentsline{toc}{subsection}{\protect
\numberline{\arabic{section}.\arabic{subsection}}{#1}} \vglue .4cm
\pagebreak[3] \noindent{\it \thesubsection.
#1}\nopagebreak[4]\par\vskip .3cm}

\renewcommand{\theequation}{\thesection.\arabic{equation}}

\newlength{\extraspace}
\newlength{\extraspaces}
\newcounter{dummy}
\newcommand{\bc}{\begin{center}}
\newcommand{\ec}{\end{center}}
\newcommand{\be}{\begin{equation}
\addtolength{\abovedisplayskip}{\extraspaces}
\addtolength{\belowdisplayskip}{\extraspaces}
\addtolength{\abovedisplayshortskip}{\extraspace}
\addtolength{\belowdisplayshortskip}{\extraspace}}
\newcommand{\ee}{\end{equation}}

\newcommand{\ba}{\begin{eqnarray}
\addtolength{\abovedisplayskip}{\extraspaces}
\addtolength{\belowdisplayskip}{\extraspaces}
\addtolength{\abovedisplayshortskip}{\extraspace}
\addtolength{\belowdisplayshortskip}{\extraspace}}
\newcommand{\ea}{\end{eqnarray}}

\newcommand{\ban}{\begin{eqnarray*}
\addtolength{\abovedisplayskip}{\extraspaces}
\addtolength{\belowdisplayskip}{\extraspaces}
\addtolength{\abovedisplayshortskip}{\extraspace}
\addtolength{\belowdisplayshortskip}{\extraspace}}
\newcommand{\ean}{\end{eqnarray*}}
\newcommand{\baa}{
\addtocounter{equation}{1} \setcounter{dummy}{\value{equation}}
\setcounter{equation}{0}
\renewcommand{\theequation}{\thesection.\arabic{dummy}\alph{equation}}
\begin{eqnarray}
\addtolength{\abovedisplayskip}{\extraspaces}
\addtolength{\belowdisplayskip}{\extraspaces}
\addtolength{\abovedisplayshortskip}{\extraspace}
\addtolength{\belowdisplayshortskip}{\extraspace}}
\newcommand{\eaa}{
\end{eqnarray}
\setcounter{equation}{\value{dummy}}
\renewcommand{\theequation}{\thesection.\arabic{equation}}}

\newcommand{\vev}[1]{\left\langle #1\right\rangle}

\newcommand{\half}{\frac{1}{2}}
\newcommand{\del}{\partial}
\newcommand{\eol}{\nonumber \\}

\newcommand{\delb}{\bar{\del}}

\newcommand{\alphap}{ \kappa }
\newcommand{\F}{{\cal F}}
\newcommand{\Q}{{\sf Q}}
\newcommand{\G}{{\sf G}}

\newcommand{\su}{{\widehat {\frak su}{ (2)}}}

\renewcommand{\L}{{\cal L}}

\begin{document}

\begin{flushright}
%June 2005\\
{\tt hep-th/0507037} \\
PUPT-2169 $\qquad$
\end{flushright}
\vspace{2cm}

\thispagestyle{empty}

\begin{center}
{\Large\bf  Minimal  $\bf{AdS_3}$
 \\[13mm] }

{\sc L.~Rastelli, M.~Wijnholt}\\[2.5mm]
{\it Physics Department, Jadwin Hall, Princeton University\\
Princeton, NJ 08544}\\
[30mm]

{\sc Abstract}

\end{center}

\noindent

We show that Type IIB string theory on $AdS_3 \times S^3 \times M_4$ with $p$ units
of NS flux contains an integrable subsector,  isomorphic
to the minimal $(p,1)$  bosonic string. To this end, we construct a  topological string theory with target space Euclidean
$AdS_3 \times S^3$.  We use a variant of
Hamiltonian reduction to prove its equivalence to
 the minimal $(p,1)$ string.
 The topological theory is then  embedded in the
  physical ten-dimensional IIB string theory.
   Correlators of tachyons in the minimal string are mapped to correlators of spacetime chiral primaries in the IIB theory, in the presence of background 5-form RR flux.
We also uncover a ground ring structure in $AdS_3 \times S^3$
analogous to the well-known ground ring of the minimal string.
This tractable model provides a literal incarnation of the idea that the
 holographic direction of AdS space  is the Liouville field.
We discuss a few generalizations, in particular we show
that the $N=4$ topological string on an $A_{p-1}$ ALE singularity
also reduces to the $(p,1)$ minimal string.

  \vfill

\newpage

\renewcommand{\Large}{\normalsize}
\tableofcontents

\newsection{Introduction and Summary}

Exactly solvable models have played an important historical
role in physics. In string theory, topological strings  and strings in low dimensions
are two related classes of models often amenable to exact treatment.
Apart from their intrinsic  value as tractable toy models,
topological strings can also be embedded in physical
superstring theories; the topological
observables correspond to a  subsector of BPS
quantities of the physical theory.  The best known examples
are Type II compactifications on Calabi-Yau threefolds,
where the topological string of the Calabi-Yau computes
certain $F$-terms of the four-dimensional effective action \cite{Bershadsky:1993cx, Antoniadis:1993ze}.

It is of great interest to extend these ideas to  backgrounds
involving an Anti-de Sitter factor,
with the prospect of obtaining new insights into the
 AdS/CFT correspondence  and the workings of open/closed duality.
 Recent results \cite{jevicki, Berenstein:2004kk, Lin:2004nb}  about  1/2 BPS configurations
 in AdS spaces and  their dual field theories
 are strongly suggestive of a topological string structure.
 In fact there is a natural class of exactly solvable models that seem tailor-made to be embedded
  into backgrounds
of the form ${ AdS} \times X$, for $X$  a compact manifold:
 the  string theories defined by coupling (minimal) $ c \leq 1$ matter to gravity.
Computations of special BPS quantities should be captured
by the topological string on ${ AdS} \times X$. Moreover
 - one is tempted to speculate -  the  AdS factor of the topological
 sigma model may be replaced by  Liouville CFT,
 while the $X$ factor may reduce to  $c \leq 1$ matter.

In this paper we show that this general guess is precisely
realized in what is, in some technical sense, the simplest AdS
background: IIB string theory on ${AdS}_3 \times S^3 \times M_4$,
with $p$ units of NS flux. This well-known background arises as
the near horizon geometry of  $Q_1$ fundamental strings and $Q_5
\equiv p$ NS5 branes wrapping the four-manifold $M_4$, which can
be either $K_3$ or $T^4$.  In this concrete example the worldsheet
description is under complete control \cite{Giveon:1998ns} and we
can carry out a very explicit analysis. We are going to prove that
the  B model with target  space (Euclidean) $AdS_3 \times S^3$ is
equivalent to the $(p,1)$ bosonic string, to all orders in the
topological genus expansion. This equivalence could be phrased as
``taking seriously'' the $SL(2)$ current algebra in the KPZ
description \cite{Knizhnik:1988ak} of worldsheet gravity for the
minimal string: the $SL(2)$ is given a literal spacetime
interpretation  as  the $AdS_3$ factor of the sigma model.
Similarly, the minimal matter is carved out of the $SU(2)$ current
algebra for the $S^3$ factor of the sigma model. As in the
Calabi-Yau case, the topological theory  can be viewed as a BPS
subsector of the physical theory. Special correlators of  chiral
primaries in the IIB theory, in the presence of background 5-form
RR flux, are thus reduced to computations in the minimal $(p,1)$
string. We hasten to add that the minimal $(p,1)$ theory
considered in this paper is not quite the same as the purely
topological theory discussed {\it e.g.} in \cite{WittenN,
Dijkgraaf:1990qw, Dijkgraaf:1991qh} and which can be realized  as
an $N=2$ minimal model coupled to topological gravity. This latter
theory only captures the ``resonant''  Liouville amplitudes of the
$(p,1)$ minimal string.

We are actually going to base our technical
analysis on the Euclidean version of $AdS_3$, the  hyperbolic space $H_3^+$,
since  the topological theory  seems more naturally defined on an Euclidean manifold.
The supersymmetric  sigma model on $H_3^+ \times S^3$ consists
of an $SU(2)_{p-2}$ WZW model for the sphere, of an $SU(2)_{-p-2}$ WZW
model for the hyperbolic space, and of  six real free fermions \cite{Giveon:1998ns}.
We define the topological string theory by the usual
procedure of twisting the $(2, 2)$  worldsheet supersymmetry.
The A model turns out to be trivial, so we focus on the B model.
Because of the special symmetries
of this theory,  four of the six coordinates are redundant -
the two ``transverse'' coordinates
along the boundary of $H_3^+$ and two of the $S^3$ coordinates.
Indeed, the structure of the  BRST operator of the topological theory is such that these
coordinates  and  four of the free fermions fill
two   ``$b c \beta \gamma$ quartets'',
which can be argued to decouple. The reduced theory is identical to the $(p,1)$
bosonic string:  The remaining $S^3$ coordinate provides a
Coulomb gas description of the $(p,1)$ matter; the ``holographic''
coordinate of $H_3^+$ becomes the Liouville field; and the
two remaining free fermions transmute into the diffeomorphism
ghosts of the bosonic string. We
demonstrate the equivalence at the full quantum level,
showing that the rules for computation of string amplitudes
inherited from the B model are the expected ones.

A precise dictionary   describes the embedding of the minimal string into IIB string theory on $H^+_3 \times S^3 \times M_4$:

\begin{itemize}
\item
The  tachyons of the minimal string map to
 universal 1/2 BPS operators in  $H^+_3 \times S^3$,
 which are   (chiral, chiral) both on the  worldsheet and in spacetime.
 By ``universal'' we mean  independent of the details of $M_4$.
 The  condition of being chiral in spacetime says
 that their angular momentum $j$ around
 a preferred   axis of the $S^3$ equals the spacetime conformal dimension $h$.
 These states can be organized in a sequence with increasing $h=j$
 quantum number, with   $h = (n-1)/2 $,  one state for each integer $n \geq 1$, $n \neq 0$ mod $p$.
 (See Figure 2 in section 4.4).
 The first $p-1$ of them are constructed in $H^+_3 \times S^3$ by combining
 conventional primaries of the two current algebras,
 and map to the ``small phase space'' of the minimal theory. The higher states are constructed
 using the operation of ``spectral flow'' - they are long string states that are supported
 on worldsheets that wrap the boundary
 of $H_3^+$ multiple times - and map to the ``gravitational descendants'' of the minimal theory.
 Every $p$-th state is missing both in the minimal string and in the IIB theory.
 This is  natural from the viewpoint of the $p$-KdV hierarchy, where
 every $p$-th flow parameter is redundant.

 \item

 Besides the tachyons, which carry ghost number one, the minimal string
 has non-trivial cohomology elements at ghost number zero. These states
 form the so-called ground ring and are the hallmark of  integrability of the model.  The
 ground ring lifts to an analogous ring structure  in the IIB theory.
  We find non-trivial cohomology classes in  $H_3^+ \times S^3 $, carrying zero ghost number with
  respect to  the ghosts of the  10d superstring. The appearance
 of the ground ring structure follows from the fact that
 the $\su_{p-2}$ current algebra representations  are degenerate:
 imitating the construction of \cite{Lian:1991gk,Bouwknegt:1991yg,Imbimbo:1991ia,Govindarajan:1992kv},
 each primitive null  over an $\su_{p-2}$ primary gives rise to
 a ground ring state; the $H_3^+$ sector provides
 the  ``gravitational dressing''.  In principle it should
 be possible to organize  calculations of 1/2 BPS amplitudes in the IIB theory in terms
 of ring multiplication rules, in analogy with the calculations
 in the minimal string \cite{Seiberg:2003nm}.
 We expect ground ring structures to be ubiquitous in supersymmetric backgrounds
 of the form $AdS \times X$, though demonstrating
 their existence from a worldsheet viewpoint as we do here
 may be possible only in special cases.

 \item

Finally, D-branes of the minimal string must correspond to supersymmetric B branes
in $H_3^+ \times S^3$.   We believe that the FZZT
brane of the minimal string lifts to the supersymmetric
$H_2^+ \times S^2$ brane discussed in \cite{Bachas:2000fr,Ponsot:2001gt}, but
we leave a detailed comparison for future work.

\end{itemize}

Our work should have implications for the $AdS_3/CFT_2$ correspondence.
The dual boundary CFT is a deformation of the
symmetric product sigma model ${\rm Sym}^{Q_1 Q_5} (M_4)$,
and is not very well understood.
It should be viewed  in some sense as the theory of the long strings
that make up the geometry before taking the near horizon limit. Thus the holographic
correspondence is not an instance of open/closed duality (that would
be the case for the S-dual D1/D5 background), and is rather more similar in spirit to matrix string theory
\cite{Dijkgraaf:1997vv}\cite{Hosomichi:1999uj}.
For amplitudes of chiral primaries we are entitled to expect simplifications.
Many details remain
to be worked out,
but we describe the main idea, that the computation of such amplitudes can be reduced to a counting
problem familiar from Hurwitz theory.

Our results are connected with several ideas of current interest.
In this paper we only begin to explore some of these relations:
\begin{itemize}
\item
An obvious connection is with the ``bubbling'' $AdS_3$ geometries
of \cite{Lunin:2002iz,Lin:2004nb,Martelli:2004xq,Liu:2004hy}.
As in these works, we are focusing on 1/2 BPS states.
Moving in the small phase space of the minimal string corresponds
to exploring  (a subset of) 1/2 BPS configurations in IIB that preserve
the $H_3^+ \times S^3$ asymptotic boundary conditions. It would be
interesting to understand in detail the six-dimensional geometric interpretation
of the small phase space and of its $W_p$ integrable structure.
\item
The background $H_3^+ \times S^3$ provides a ``critical'' ($\hat c= 3$) topological
realization of the $(p,1)$ theories.
Recently another critical realization of the $(p,1)$
models has been proposed, as the B model on
a target Calabi-Yau  related to the ground ring curve \cite{Aganagic:2003qj}.
We conjecture that the $H_3^+ \times S^3$ sigma
model is T-dual to this Calabi-Yau, in analogy
with the T-duality between  NS5 brane geometries
and ALE spaces \cite{Ooguri:1995wj}.
 \item
 The B model on  $H_3^+ \times S^3$ is arguably the simplest critical topological
 theory  with torsion. It should provide a good testing
 ground for the abstract framework of twisted generalized complex geometry \cite{Hitchin,Gualtieri}.
  \item
Besides the $AdS_3/CFT_2$ correspondence,
we may also discuss open/closed duality.
To this end we need to introduce D-branes in the $H_3^+ \times S^3$
background. The open string field theory on the FZZT
branes of the $(p,1)$ model  reduces to a Kontsevich matrix model
\cite{Gaiotto:2003yb, Maldacena:2004sn, Hashimoto:2005bf,Ghoshal:2004mw,Giusto:2004mt}, while
the open string field theory on infinitely many decayed ZZ branes corresponds
to the doubled scaled matrix model \cite{McGreevy:2003kb,Klebanov:2003km}.
It would be interesting to lift these statements to the superstring theory on $H_3^+ \times S^3 \times M_4$.
\end{itemize}

\begin{figure}[t]
\begin{center}
\leavevmode\hbox{\epsfxsize=5.5in
\epsffile{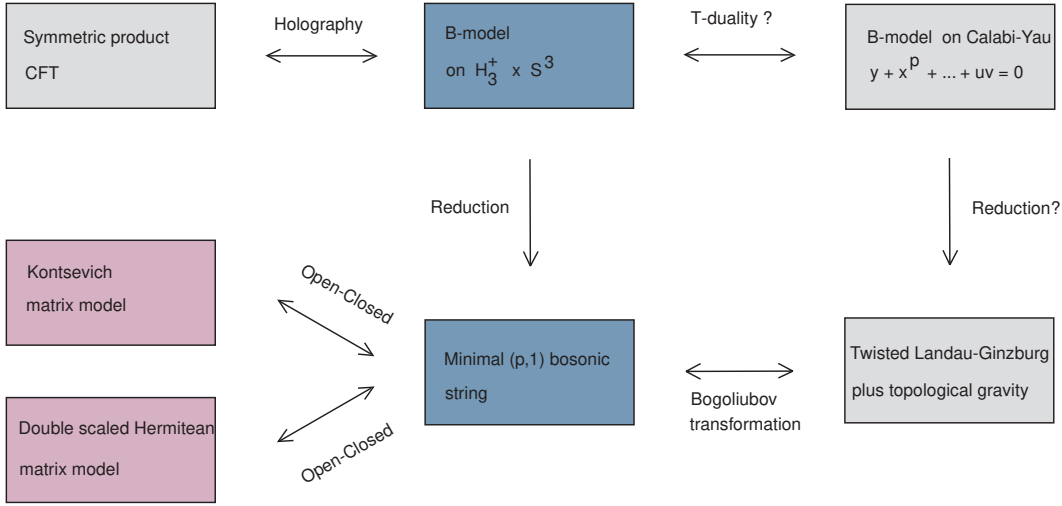}}\\[7mm]
\caption{ \it
 Web of relations centering on the  $(p,1)$ model coupled to gravity. In this paper
we focus on the two central (blue) boxes.}
\end{center}
\end{figure}

One can envision  several generalizations of this work. In this paper we focus
on the purely Neveu-Schwarz  $H_3^+ \times S^3$ background (apart from the RR ``graviphoton''
5-form flux needed to twist the theory). It is possible to turn
on additional self-dual 3-form  RR flux in six dimensions while preserving
supersymmetry. An obvious question is whether this more
general background can be reduced to a minimal string.
A simple speculation is that the 3-form RR deformation corresponds
to deforming the KdV hierarchy into the mKdV hierarchy
of complex matrix models.  This amounts to embedding
the minimal bosonic theory into the 0A theory \cite{Klebanov:2003wg,Johnson:2003hy,
Seiberg:2004ei}.
If this guess is correct, the self-dual 3-form
 RR flux in six dimensions would map
to the RR background flux of the 0A theory. The natural framework
to investigate this issue is the hybrid formalism of
\cite{Berkovits:1999im}. Other interesting generalizations are to
supersymmetric backgrounds of the form $H_3^+ \times {\cal N}$,
where ${\cal N}$ is a coset model. We briefly study the simple
example  $H_3^+ \times S^1$ and show that it reduces to the $c=1$
bosonic string at the self-dual radius. In the final section we
consider the  $N=4$ topological
 string on an $A_{p-1}$ ALE singularity, and show that it
 can also be mapped to the $(p,1)$
 minimal string by the same reasoning used
 for $AdS_3 \times S^3$.

The organization of the paper is as follows. In section 2 the topological
theory on $H_3^+ \times S^3$ is defined. In section 3 we prove that this theory reduces
to the minimal string. In section 4 we describe its embedding
 into the physical theory. Section 5 contains some  remarks
 on the holographic correspondence. In section 6  we comment
 on the relation to Calabi-Yau spaces and discuss some
 generalizations.
 Appendix A contains some mathematical details
 needed in the reduction of section 3.
In Appendix B we review the $(p,q)$ minimal theories
and spell out our viewpoint about the $(p,1)$ models, which require
 special consideration.

Topological strings on $AdS_3 \times {\cal N}$ have previously been considered in
\cite{Sugawara:1999fq}.

\newsection{Topological string theory on $H_3^+ \times S^3$ }

In this section we construct a topological string theory with target
space $H_3^+ \times S^3$.  We include a review of the worldsheet
 CFT and of its supersymmetry structure.\footnote{The Mathematica package OPEdefs.m
\cite{Thielemans:1991uw} was used to check some OPEs in this paper.}

\newsubsection{RNS description}

We start with a brief summary of the RNS worldsheet description of superstring theory
with target space $H_3^+ \times S^3$ \cite{Giveon:1998ns}.
We  will fix the amount of NS flux $H$ for the remainder as
\be
\int_{S^3} H =  p \;.
\ee
The bosonic sector of the CFT consists of two WZW models, one for the sphere and one for the
hyperbolic space.
The $S^3$ factor is described by an
 $\su$ current algebra at level $ k=p-2$,
 with central charge $3k/(k+2) = 3 - 6/p$.
  We denote the left-moving $\su_{p-2}$ generators by $j_\pm = j_1 \pm i j_2$ and $ j_3$, with OPEs
\ba
   j_+(z) \cdot j_-(w) & \sim &{p-2\over (z-w)^2} +
{2 \, j_3(w) \over z-w}  \eol
  j_3(z) \cdot j_\pm(w) & \sim & \pm {j_\pm(w) \over z-w} \eol
  j_3(z) \cdot j_3(w) & \sim & {p-2 \over 2 (z-w)^2} \,  .
\ea
Similarly we have right-moving generators $\tilde j_\pm (\bar z)$, $\tilde j_3 (\bar z)$.
The hyperbolic space $H_3^+ $ is described by an $ SL(2,{\bf C})/SU(2)$ coset
model at level $p+2$,  equivalently by
 an $\su$ current algebra at level $k'=-p-2$,
with central charge $3k'/(k'+2) = 3 + 6/p$. The total bosonic central
charge of $S^3 \times H_3^+$ is then six, as expected.
  We denote
the left-moving $\su_{-p-2}$ generators by  $k_\pm = k_1 \pm i k_2$ and $ k_3$, with OPEs
\begin{eqnarray}
   k_+(z) \cdot k_-(w) & \sim & - {p+2 \over (z-w)^2} +
{2\,k_3(w) \over z-w}  \eol
  k_3(z) \cdot k_\pm(w) & \sim & \pm {k_\pm(w) \over z-w} \eol
  k_3(z) \cdot k_3(w) & \sim & -{p +2 \over2  (z-w)^2} \, .
\end{eqnarray}
Similarly for the right-moving generators $\tilde k_\pm (\bar z)$, $\tilde k_3 (\bar z)$.
It is convenient to adopt a six-dimensional notation
and introduce the symbol $J_a$ with $a=1, \dots 6$ to denote
all the (left-moving)  currents:
\ba
J_a & \equiv & j_a \qquad{\rm  for} \;  a=1,2,3 \nonumber \\
J_a & \equiv&  k_{a-3}  \quad{\rm  for} \;  a=4,5,6 \,.
\ea
The fermionic sector of the CFT consists
of the superpartners of the currents, six free  fermions  $\lambda^a (z)$, $a = 1, .. ,6$.
(Similarly we have $\tilde J_a (\bar z)$ and $\tilde \lambda^a(\bar z)$. In the following
we shall usually avoid  explicit mention of the right-movers.)
Notice that the currents have lower indices, and the fermions
have upper indices. This is a natural notation since
the (zero modes of the) currents correspond to tangent vectors, while the
(zero modes of the) fermions  correspond to one-forms.

\smallskip

Let us introduce a metric $h_{ab}$ in this six-dimensional vector space:
\be\label{hmetric}
h_{ab} = \half \left(
\matrix{ p  \,{\bf 1}_{3 \times 3} && 0 \cr
  0 && - p \,{\bf 1}_{3 \times 3}  } \right)_{ab}
\ee
The $h_{ab}$ are the residues of the double poles of the currents,
up to a shift by $\delta_{ab}$ (= half the dual Coxeter number of ${\frak su(2)}$).
The fermions are normalized to satisfy the following OPEs,
\be
\lambda^a(z) \cdot \lambda^b(w) \sim {h^{ab} \over z-w}  \, .
\ee
In these notations,  the stress tensor and supercurrent are given
by the simple standard expressions
\ba
{\sf G} &=& J_a \lambda^a -{1\over 6} h_{cd} f^d_{ab} \lambda^a \lambda^b \lambda^c\, , \eol
{\sf T} &=& \half(h^{ab} J_a J_b + h_{ab} \del \lambda^a \lambda^b) \, .
\ea
Here $f_{ab}^c $ are the structure constants of ${\frak su(2)} \otimes {\frak su(2)}$.
It is easy to check that
\be
{\sf G}(z) \cdot {\sf G}(w) \sim {6 \over (z-w)^3} + {2{\sf T}\over z-w} \,.
\ee
In particular the central charge is $\hat{c} = c/3 = 3$.

\newsubsection{$(2,2)$ structure}

In order to define a topological string theory we need to
identify an extended $(2,2)$ worldsheet supersymmetry.
For the case at hand, there is a standard construction that
we now briefly review  (\cite{Spindel:1988sr}, see also
\cite{Parkhomenko:1992dq,Getzler:1993py})\footnote{Readers willing to take
on faith the basic definitions (\ref{cs}) can skip directly there.}.

The conditions for extended susy for a general $\sigma$-model with
torsion $H $ are well-known \cite{Gates:1984nk}. The  target space manifold must admit
 two integrable complex structures ${\cal J}_{\pm}$;
  the metric must be Hermitian with respect to both complex structures;
  and ${\cal J}_\pm$ must be covariantly constant, each
  with respect  to a different affine connection,
  \ba
  \Gamma^{\pm \mu}_{\rho \nu}  \equiv \Gamma^\mu_{\rho \nu} \pm g^{\mu \sigma} H_{\sigma \rho \nu} \\
\nabla^\pm_\rho  {\cal J}^\mu_{\pm \nu}  =   0 \nonumber \,.
\ea
 In the case of group manifolds, there is a canonical way to construct
 globally defined almost complex structures ${\cal J}^\mu_{\pm \nu}$:
 start  with a constant matrix ${\cal J}^a_{\; b}$
 acting on the Lie algebra, and use the left and right group
 action to obtain respectively  ${\cal J}^\mu_{- \nu}$ and  ${\cal J}^\mu_{+\nu}$.
 The various conditions on ${\cal J}^\mu_{\pm \nu}$
 translate into conditions for   ${\cal J}^a_{\; b}$.

  Let us see in more detail  how this  works. The
  left- and right-invariant global vector fields   $J_a$ and $\tilde J_a$ on the group manifold are
  covariantly constant  with respect to the two connections,\footnote{By a slight abuse of notation,
 here we use the symbols $J_a$, $\tilde J_a$  to denote the zero modes
  of the current $J_a(z)$, $\tilde J_a (\bar z)$}
 \be
 \nabla^- J_a = 0 \, , \quad \nabla^+ \tilde J_a = 0 \,.
 \ee
 Given a constant matrix  ${\cal J}^a_{\;b}$ acting on the Lie algebra
 and squaring to minus one,
 we  parallel-transport it in two different ways to give
the two candidate complex structures ${\cal J}^\mu_{\pm \nu}$, globally defined
on the manifold,
\be
{\cal J}^\mu_{- \nu} =J^\mu_a {\cal J}^a_{\; b} J^b_\nu \, , \quad
{\cal J}^\mu_{+ \nu} =\tilde J^\mu_a {\cal J}^a_{\; b} \tilde  J^b_\nu \,.
\ee
The point of this construction is that the resulting  ${\cal J}^\mu_{\nu}$ are automatically covariantly
constant with respect to $\Gamma^\pm$. The hermiticity condition
for $g_{\mu \nu}$ with respect to
 ${\cal J}^\mu_{\pm \nu}$ translates into the hermiticity condition of
the metric $h_{ab}$ on the Lie algebra with respect
to ${\cal J}^a_b$. To express the integrability condition,
it is convenient to use ${\cal J}^a_b$ to split  the real indices $a, b, \dots$ into
holomorphic  indices $i,j,\ldots$ and  antiholomorphic indices $\bar i , \bar j, \ldots$.
Then the  integrability  of  ${\cal J}^\mu_{\pm \nu}$  translates
into the condition $f_{ij}^{\bar k} = f_{\bar i \bar j}^k = 0$,  that is
 holomorphic and antiholomorphic generators form closed subalgebras.
In summary, the existence of $(2,2)$ supersymmetry on an even
dimensional group manifold is associated to a decomposition of the
complexified Lie algebra ${\frak g}^{{c}} = {\frak g}_- \oplus
{\frak g}_+$, such that ${\frak g}_{\pm}$ (the $\pm \sqrt{-1} $
eigenspaces of ${\cal J}^a_{\; b}$) are subalgebras and are
maximally isotropic with respect to  the  metric $h_{ab}$ (this
means $h_{ij } = h_{\bar i \bar j} = 0$). The structure $({\frak
g}^c, {\frak g}_-, {\frak g}_+)$ is called a Manin triple.

Finally we are in the position to quote the general expressions
for the  $N=2$ supercurrents:
\ba \label{N2}
{\sf G}^+ &=& J_i \lambda^i - {1\over 6} f_{ij}^k \lambda^i \lambda^j \lambda_k  \, ,\eol
{\sf G}^- &=& J^i \lambda_i - {1\over 6} f^{ij}_k \lambda_i \lambda_j \lambda^k \,.
\ea
Here we have used the Hermitian metric $h_{i\bar{j}}$ to get rid of all anti-holomorphic indices by raising
or lowering them. The decomposition of ${\sf G} = {\sf G}^+ + {\sf G}^-$ hinges on the condition
$f_{ijk} = f_{\bar i \bar j \bar k} = 0$, that is, on the integrability of the complex structures.

Let us now specify the choice of complex structure in our concrete example.
 We just need to indicate how the  the real Lie algebra indices $a,b = 1, \cdots 6$ split into
 holomorphic indices $i,j=1,2,3$ and  and anti-holomorphic $\bar{i},\bar{j} = 1,2,3$. We choose
\ba \label{cs}
J_{z_1} = j_-  \qquad & J_{z_2}  = +  k_+   &  \qquad J_{z_3} = j_3 +  k_3  \eol
 J_{\bar{z}_1} = j_+    \qquad & J_{\bar{z}_2}   =  -k_-  &  \qquad  J_{\bar{z}_3} = j_3 -k_3 \, .
\ea
It is easy to check that this defines a Manin triple. This choice
of complex structure is essentially unique up to symmetries.  For
example we could apply the automorphisms $k_+ \leftrightarrow k_-,
k_3 \leftrightarrow -k_3$, or  $j_+ \leftrightarrow j_-, j_3
\leftrightarrow -j_3$, etc., which correspond to spacetime
isometries. We make the choice (\ref{cs}) for future convenience,
since  the $(p,1)$ string will arise with a minimum amount of
field redefinitions. The fermions with holomorphic  and
anti-holomorphic indices are defined as \ba \lambda^{z_1} =\half
(\lambda^1 + i \lambda^2) \qquad   \lambda^{z_2} =+\half
(\lambda^4 - i \lambda^5) \qquad \lambda^{z_3} =\half (\lambda^3 +
\lambda^6) \, , \eol \lambda^{\bar{z}_1} =\half (\lambda^1 - i
\lambda^2) \qquad   \lambda^{\bar{z}_2} =-\half (\lambda^4 + i
\lambda^5) \qquad \lambda^{\bar{z}_3} =\half (\lambda^3 -
\lambda^6) \,. \ea
The metric on the Lie algebra takes the form
\be
h_{i\bar{j}} = p\, \delta_{i\bar{j}} \, , \quad  h_{i j} = h_{\bar i \bar j} = 0 \, ,
\ee
which is manifestly Hermitian.
Specializing (\ref{N2}) to our  case we obtain the following $N=2$ structure
\begin{eqnarray} \label{N2expl}
\sf{T} &=&
 {1\over 2p}(2 j_3 j_3 + j_+ j_- + j_- j_+ - 2 k_3 k_3 - k_- k_+ - k_+ k_-) \eol
& &+ {1\over 2}\left(\del c^i b_i - c^i \del b_i \right)
\eol
\sf{G}^+ &=&c^1(j_-) +  c^2( k_+) + c^3( j_3 +  k_3 + c^1 b_1 - c^2 b_2 )
   \eol
\sf{G}^- &=& {1\over p}\left( b_1 j_+
-b_2  k_- + b_3( j_3 -  k_3 + c^2 b_2 + c^1 b_1)
 \right)  \eol
\sf{J} &=& c^i b_i  -{2 \over p} J_3^{\rm tot}
\end{eqnarray}
with
\be
  J_3^{\rm tot} \equiv j_3 +  k_3+ c^1 b_1- c^2 b_2   = \{ {{\sf G}}^+_0, b_3 \} .
\ee
We have renamed the fermions as $c^i \equiv \lambda^i$ and $b_i \equiv \lambda_i$,
which will be a useful notation. From now on we shall never
lower or raise indices: $b_i$ $(c^i)$ will consistently have lower (upper) indices.
They satisfy the OPEs
\be
b_i(z) \; c^j(w) \sim {\delta_i^j \over z-w} .
\ee
In fact there is a whole  family of $N=2$ structures generalizing (\ref{N2expl}),
\begin{eqnarray}
\label{N2explalpha}
\sf{T} &=&
 {1\over 2p}(2 j_3 j_3 + j_+ j_- + j_- j_+ - 2 k_3 k_3 - k_- k_+ - k_+ k_-) \eol
& &+ {1\over 2}\left(\del c^i b_i - c^i \del b_i \right)
+ {\alphap \over 2p}\del J_3^{\rm tot}
\eol
\sf{G}^+ &=&c^1(j_-) +  c^2( k_+) + c^3( j_3 +  k_3 + c^1 b_1 - c^2 b_2 )
   \eol
\sf{G}^- &=& {1\over p}\left( b_1 j_+
-b_2  k_- + b_3( j_3 -  k_3 + c^2 b_2 + c^1 b_1)
 + \alphap \del b_3 \right)  \eol
\sf{J} &=& c^i b_i  + {\alphap-2 \over p} J_3^{\rm tot}\,.
\end{eqnarray}
A preferred value of the parameter $\alphap$ will emerge when we discuss the reduction to the minimal
string, but it will be useful to maintain the explicit $\alphap$ dependence.

\newsubsection{Twisting}

The next step is to twist the $(2,2)$ structure \cite{Witten:1991zz}.
There are four choices. We can twist ${\sf T}$ either to ${\sf T} + \half \del {\sf J}$ or
to ${\sf T} - \half \del {\sf J}$, and similarly for the right-movers. We will
focus on the B model with $(+, +)$ twist,
which will  be shown to be equivalent to the minimal string. There is no obstruction
in defining the B model because the worldsheet fermions are free, and the axial
current is manifestly non-anomalous. The A model will turn out to be trivial.
We hasten to add that because of the torsion, one's usual geometric intuition about
observables in the A and B models does not readily apply. In particular
B model amplitudes generically receive quantum corrections \cite{Kapustin:2004gv}.

\smallskip

Using this topological sigma model we define
a  string theory by ``coupling it to topological gravity''. In practice one can  ignore
the topological gravity multiplet and compute amplitudes using
an analogy with the bosonic string, as we now schematically review.
The analogy  is
\begin{eqnarray}\label{bosanalogy}
  {\sf T} + \half \del{\sf  J }  &  \to & T \eol
{\sf G}^+    & \to   & J_{\rm BRST}     \eol
   {\sf G}^-  & \to & B \eol
{\sf J} & \to &   J_{\rm ghost} \,
\end{eqnarray}
Here $T$, $J_{BRST}$, $B$ and $J_{ghost}$ are the usual stress tensor, BRST current,
antighost and ghost number current of the bosonic string.
It is possible to add  improvement terms to $j_{BRST}$ and $J_{ghost}$
such that
$T$, $j_{BRST}$, $B$ and $J_{ghost}$  generate {\it precisely} a twisted $N=2$ algebra \cite{Bershadsky:1992pe}.

The physical states of the topological
theory  are cohomology classes of ${\sf G}_0^+$,
that is, chiral primaries of the worldsheet $N=2$ algebra.
If the  $N=2$ algebra is unitary, chiral primaries
are automatically annihilated by ${\sf G}_0^-$. In our case
the $N=2$ algebra is not unitary.
 As in the bosonic string,
proper care must be taken
  in combining left- and right-movers to
 form closed string states: the closed string Hilbert space
 is defined as the semi-relative cohomology of
 of ${\sf Q} \equiv {\sf G}^+_0 + \bar  {\sf G}^+_0 $,
which is the ${\sf Q}$-cohomology on the complex of states annihilated
by ${\sf G}_0^- - \bar {\sf G}_0^- $,
\be
 {\cal H}_{\rm Hilbert} = H_{\sf Q}({\rm  Ker}_{{\sf G}_0^- - \bar {\sf G}_0^-}) \,.
\ee
The measure on the moduli space of Riemann surfaces
is given by insertions of ${\sf G}^-$ contracted with quadratic differentials,
\be
{\sf G}^-(\mu_k) = \int d^2 z \,  {\sf G}^-_{zz}\,  \mu_{k \,\bar{z}}^z  \,.
\ee
Using the substitution   (\ref{bosanalogy}) one defines string amplitudes
for genus $g>1$ through the formula
\begin{equation}\label{}
    F_g(t) = \left<  \prod_{j=1}^{3g-3} |{\sf G}^-(\mu_j)|^2 \  \
    e^{\sum_i t_i \int d^2 z \, {\sf G}^-_{-1} \bar{ \sf G}^-_{-1} \phi^i  }\right> \,.
\end{equation}
Here $\{ \phi_i \}$ are (unintegrated) physical vertex operators,
and   absolute values squared  denote contributions from left- and
right-movers.
For genus zero there is a similar formula, except that we must
 absorb  Killing vectors fixing the position of three vertex operators,
\begin{equation}\label{}
  \del_{i} \del_{j} \del_{k} F_0(t)
  = \left< \phi^{i} \phi^j \phi^k
   e^{\sum_l t_l \int d^2 z\,  {\sf G}^-_{-1} \bar{ \sf G}^-_{-1} \phi^l }\right> \,.
\end{equation}
Finally the genus one amplitude is often presented in the operator formalism,
\begin{equation}\label{}
   F_1  = \half \int {d^2 \tau \over {\rm Im}(\tau)}
   {\rm Tr}\left[ (-1)^{f_L+f_R} f_L f_R\, q^{{\sf L}_0 - \half{\sf  J}_0} \bar{q}^{\bar{\sf L}_0- \half \bar{\sf J}_0} \right]\, ,
\end{equation}
where $q = e^{2\pi i \tau}$ and $f$ is the worldsheet fermion number.

The analogy (\ref{bosanalogy}) will be an important guiding principle in
relating our topological theory to the minimal string.

\newsection{Reduction to the minimal string}

We are now going to argue that the topological
string theory on $H_3^+ \times S^3$ is equivalent to the minimal $(p,1)$ string theory.
To this end, it is useful to use a free field representation for $H^+_3 \times S^3$.

\newsubsection{Wakimoto representation}

There is a well-known free field realization of  the $\su$ current algebra
 in terms of a linear dilaton and of a
first-order $(\beta, \gamma)$ system of dimensions $(1,0)$.
All in all, we introduce two $(\beta, \gamma)$ systems
and two linear dilaton fields $\varphi$ and $x$ with central charges $c_{\varphi} = 1+ 6/p$
and $c_x = 1 - 6/p$. The OPEs read (notice that we take $\alpha'=1$)
\ba
\beta_L(z) \gamma_L(w) \sim -{1\over z-w}, & \qquad \varphi(z) \varphi(w) \sim - \frac{1}{2} \log (z-w),  \eol
\beta_M(z) \gamma_M(w) \sim -{1\over z-w}, & \qquad x(z) x(w) \sim - \frac{1}{2} \log (z-w) \, .
\ea
For $S^3$ we represent the $\su_{p-2}$ current algebra as
\begin{eqnarray}
 j_+ &=& \beta_M \eol
j_3 &=& + \gamma_M \beta_M + i \sqrt{p}\del x \eol
j_- &=& - \gamma^2_M \beta_M -i  2 \sqrt{p}  \gamma_M\del x - (p-2) \del \gamma_M \,.
\end{eqnarray}
Similarly for $H_3^+$  we write the $\su_{-p-2}$ algebra as
\begin{eqnarray}
 k_+ &=& \beta_L \eol
k_3 &=&  \gamma_L \beta_L - \sqrt{p}\del \varphi \eol
k_- &=&
- \gamma^2_L \beta_L + 2\sqrt{p} \gamma_L \del \varphi +(p+2)
\del \gamma_L \,.
\end{eqnarray}
As we will see in more detail in section 4, the variables $\{ e^{\varphi}, \gamma_L, \bar{\gamma}_L \}$
are closely related to the Poincar\'e coordinates for $H_3^+$.
Substituting in the (untwisted) stress tensor  (\ref{N2expl}) yields
\begin{eqnarray}\label{}
 \sf{T}_j + \sf{T}_k &=&
 {1\over 2p}(2 j_3 j_3 + j_+ j_- + j_- j_+ - 2 k_3 k_3 - k_- k_+ - k_+ k_-) \eol
&=& - (\del x)^2 -{i\over \sqrt{p}} \del^2 x -(\del\varphi)^2
- {1\over \sqrt{ p}} \del^2 \varphi \eol
 & & -  \beta_M \del \gamma_M - \beta_L \del \gamma_L \,.
\end{eqnarray}
The free field representation must be supplemented with appropriate
screening charges. For $S^3$ the screeners are
\ba
\Q^M_- &=& \oint dz \, \beta_M \, e^{-2i x/\sqrt{p}}  \eol
\Q^M_+ &=& \oint dz \, \beta_M^{-p}\, e^{+2i \sqrt{p}  x}  \, ,
\ea
while for $H_3^+$,
\ba
\Q^L_- &=& \oint dz \,  \beta_L \, e^{-2 \varphi/\sqrt{p}}  \eol
\Q^L_+ &=& \oint dz \,  \beta_L^{p}\, e^{-2 \sqrt{p}  \varphi} \,.
\ea
Let us briefly review  the free field resolution \cite{Bernard:1989iy} of
the irreducible $\su$ modules, focusing on the $S^3$ factor.
We introduce the Fock spaces $\F_{m, n}$, obtained
by acting with oscillators on the vacuum of $x$-momentum $p_x = \frac{m-1}{2\sqrt{p}} + \frac{(1-n) \sqrt{p}}{2}$,
\be
\F_{m, n} \equiv {\rm Span} \{ \beta_{-i_1}  \cdots  \beta_{-i_\alpha} \,  \gamma_{-j_1} \cdots \gamma_{-j_\beta} \, a_{-k_1}  \cdots  a_{-k_\gamma}
  \, e^{ \frac{2 i ( m -1)+  2 i (1-n) p}{2 \sqrt{p} }x  } | 0 \rangle  \, \} \, .
\ee
Here the $a_n$'s are the usual oscillators for the field $x$, $i \partial x (z)= \sum_n a_n z^{-n+1}$,
and $| 0 \rangle$ is the $SL(2)$ invariant vacuum.
 Irreducible  representations of $\su_{p-2}$
 are labeled by a semi-integer spin $j$, with $0 \leq j \leq p/2 -1$. For
 a given $j$ in this range, consider the sequence
\be \label{seqSU}
  \dots   \stackrel{\Q_-^{2j+1}}{\longrightarrow}  \F_{2p- 2j-1, 1}  \stackrel{\Q_-^{p-2j-1}}{\longrightarrow}
  \F_{2j+1, 1} \stackrel{\Q_-^{2j+1}}{\longrightarrow}
  \F_{-2j-1, 1}   \stackrel{\Q_-^{p-2j-1}}{\longrightarrow}    \dots
\ee
This defines a complex, {\it i.e.}  $(\Q^M_F)^2 = 0$, where the symbol
$\Q^M_F$ denotes ${(\Q^M_-)^{2j+1}}$ or ${(\Q^M_-)^{p-2j-1}}$
according to which space it acts on. Moreover
 the sequence is exact except at the middle Fock space $\F_{2j +1, 1}$, where
the cohomology $H_{\Q^M_F}( \F_{2j +1})$  is isomorphic to the irreducible
$\su_{p-2}$ module of spin $j$ \cite{Bernard:1989iy}.
To obtain each spin $j$ representation
 with $0 \leq j \leq p/2 -1$ once,
 we consider the cohomology $H_{\Q^M_F}( \F)$, where $\F$ is the direct sum of Fock spaces
\be \label{lattice}
\F = \bigoplus_{j = 0}^{p/2 -1} \F_{[j]} \equiv \bigoplus_{j=0}^{p/2-1}\, \bigoplus_{k \in \mathbb{Z}}
\F_{\pm (2j+1) + 2k p, 1} = \bigoplus_{  m \in \mathbb{ Z}  }^{ m \neq 0 \, {\rm mod } \; p} \F_{m, 1} \, .
\ee
The absence of the Fock spaces $\F_{m= k p, 1}$  will play a role in the following.

There is a similar contruction for $\su_{-p-2}$,
with the analogous Felder BRST charge $\Q_F^L$ built
from powers of $\Q_-^L$. The representations
of $\su_{-p-2}$ that will be relevant for us are the principal discrete representations
${\cal D}_{j'}^\pm$, also labeled by  semi-integer spins $j'$.
For the purposes of this section it will be sufficient to keep track of
the representations for $S^3$.

\newsubsection{Preview}

Consider now the {\it twisted} stress tensor
\begin{eqnarray} \label{twistedT}
{\sf T}_{\rm }+ \half \del {\sf J} &=&
- (\del x)^2 + i{\alphap -2 \over \sqrt{p}} \del^2 x
- (\del\varphi)^2 - {\alphap \over \sqrt{ p}} \del^2 \varphi \eol
 & & - b_1\del c^1 + \frac{\alphap- 1}{p}  \del ( c^1 b_1) -  b_2 \del  c^2 - \frac{\alphap-1}{p} \del ( c^2 b_2)  -b_3 \del c^3 \\
 &&  - \beta_M \del \gamma_M + \frac{\alphap - 1}{p} \del(\gamma_M \beta_M) -
  \beta_L \del \gamma_L + \frac{\alphap-1}{p} \del(\gamma_L \beta_L) \,.\nonumber
\end{eqnarray}
We summarize in Table 1 the conformal dimensions of the fields and vertex operators.
For the special value $\kappa = p+ 1$,  the  additional twist
by $J_3^{\rm tot}$ is precisely of one unit,
\be \label{TT}
{\sf T}_{\rm twisted} = {\sf T}_{\rm phys} + \frac{1}{2} \partial (c^i b_i) + \partial J_3^{\rm tot} \, ,
\ee
and we see from (\ref{twistedT})
that the central charges of $x$ and $\varphi$ are the expected ones for the $(p,1)$ model coupled to gravity:
\begin{equation}
c_x = 1 - 6 {(p - 1)^2\over p}, \qquad c_\varphi = 1 + 6{(p + 1)^2\over p} \,.
\end{equation}
Moreover, the $(b_1,c^1)$ system has dimensions $(2,-1)$ and can be
identified with the diffeomorphism ghosts of the minimal string.
The remaining degrees of freedom are two pairs of $\beta \gamma$, $b c$ systems:
$(\beta_L, \gamma_L)$  $(b_2, c^2)$,  and $(\beta_M, \gamma_M)$  $(c^3, b_3)$,
of dimensions $(0, 1)  (0, 1)$.  We expect them to decouple
by the  quartet mechanism (bosonic and fermionic degrees of freedom
with the same quantum numbers canceling pairwise in the path integral).
In fact there are well-known procedures to obtain
minimal matter and Liouville CFT from $SL(2)$ WZW models, known respectively
as Hamiltonian \cite{Bershadsky:1989mf}
and KPZ \cite{Knizhnik:1988ak} reduction,
which also exploit the idea of quartet confinement.

\begin{table}
\label{dimensions}
%\begin{center}
\begin{tabular}{|c|c|c|c|c|c|c|c|c|c|c|c|c|}
\hline
  & $\beta_M$ & $\gamma_M$ & $\beta_L$ &$\gamma_L$ & $c^1$ & $c^2$ & $c^3$ & $b_1$ & $b_2$ & $b_3$  \\
\hline
$ \Delta$ & $ \frac{p+1-\alphap}{p}$ & $\frac{\alphap-1}{p}$ & $\frac{p+1-\alphap}{p} $ & $\frac{\alphap-1}{p}$  & $\frac{1-\alphap}{p}$ & $\frac{\alphap-1}{p}$ & $0$ & $\frac{\alphap+p-1}{p}$  & $\frac{ p+1-\alphap}{p}$ & $ 1 $  \\
\hline
\end{tabular}
%\end{center}
\newline
\newline

\begin{tabular}{|c|c|c|}
\hline
  & $e^{2 i \alpha x}$ & $e^{2  \beta \varphi }$ \\
  \hline
$\Delta$ & $\alpha (\alpha - \frac{\alphap - 2}{\sqrt{p}} ) $ & $-\beta (\beta + \frac{\alphap}{\sqrt{p}} ) $ \\
\hline
\end{tabular}
%\end{center}
\caption{ \it Conformal dimensions in the twisted theory, for arbitrary $\kappa$.}
\end{table}
\vskip 1mm

While this is promising, we should not
dictate any additional rules; the
equivalence with the
 minimal string should arise from the established rules for perturbative
(topological) string theory that we reviewed above.
To prove this,  we will decompose the BRST operator of the topological
theory as ${\sf G}^+_0 = {\sf Q}_1 + {\sf Q}_R$,
and show that taking the cohomology with respect to
 ${\sf Q}_R$ implements a version
of  Hamiltonian  + KPZ reduction,
reducing the field content to that of the minimal string.
Then we will find a similarity transformation that maps the $N=2$ generators
of the topological string  to the corresponding generators of the bosonic string
(\ref{bosanalogy}), up to ${\sf Q}_R$-trivial terms. In particular
the operator $\Q_1$ is mapped to the usual BRST operator
$\Q_{\rm Vir}$ of the bosonic string. Schematically, we are going to
show the following equivalences of cohomologies
\begin{equation}\label{}
    H_{{\sf G}^+_0}( H_3^+ \times S^3) \cong H_{\Q_1}( H_{{\sf Q}_R } (  H_3^+ \times S^3) )
    \cong
    H_{{\sf Q}_{\rm Vir}} ( {\rm (p,1) + gravity})\,.
\end{equation}
Finally, through the use of the similarity transformation
 the computations of topological amplitudes  are exactly mapped to the
 corresponding computations
 in the $(p,1)$ bosonic string.

We should mention at the outset that there is at least a superficial
 resemblance of this story to computations done for $G/G$ WZW models
with $G= SU(2)_{p-2}$,  where the spectrum also exactly reproduces that
of the $(p,1)$ minimal bosonic string \cite{Hu:1992xz,Aharony:1993su,Sadov:1992tf}.
Our BRST operator differs from the one used in the $G/G$ context.
There is also the crucial conceptual
difference that to compute correlation functions in the $G/G$ model one would
naturally integrate over the moduli space of flat $SU(2)$ connections, whereas in our case
we want to have a string theory and so we  integrate over the moduli
space of Riemann surfaces. We do not exclude however that a deeper
connection may be found.
Although both the interpretation and several details are different,
we found it useful to borrow some technical aspects of the work
by Sadov \cite{Sadov:1993nt}, who performed
a  cohomological analysis in the context of $G/G$ models
(see also  E.Frenkel, appendix in \cite{Mukhi:1993zb}).

\newsubsection{Setup for the reduction}

Our task is to study the cohomology of ${\sf G}^+_0$ acting
on the state space of the $H_3^+ \times S^3$ sigma model.
Using the free field representation, we are instructed
to first evaluate the cohomology of the Felder
charges on the free field state space, and then the ${\sf G}^+$ cohomology,
\be \label{problem}
H_{{\sf G}^+}(   H_3^+ \times S^3) \cong H_{{\sf G}^+}(  H_{\Q_F^M + \Q_F^L} (  \F_{}  )) \, .
\ee
Here $\F$ is the total space state of the {\it free} fields
\be
\F = \F_x \otimes \F_{\beta^M \gamma^M}  \otimes
\F_\varphi \otimes \F_{\beta^L \gamma^L}  \otimes
\F_{b_i c^i} \,.
\ee
The main idea is to split up the BRST differential as
\be
{\sf G}^+_0 = {\sf Q}_1 + {\sf Q}_2 + {\sf Q}_3, \qquad \{{\sf Q}_i , {\sf Q}_j \} =0.
\ee
where we choose
\ba
{\sf Q}_1 &=& \oint c^1 j_- \\
{\sf Q}_2 &=& \oint c^2 \beta_L  \eol
{\sf Q}_3 &=&
\oint c^3(- \sqrt{ p} \del \varphi + i \sqrt{p} \del x + \gamma_M \beta_M + c^1 b_1)
+ \oint c^3(\gamma_L \beta_L - c^2 b_2) \, , \nonumber
\ea
In relation to the previous section,
$\Q_R \equiv
\Q_2 + \Q_3$ is the operator that implements the reduction,
while $ \Q_1$ will turn out to be equivalent to $\Q_{\rm Vir}$ of the bosonic string.
 We can assign
gradings to the fields by defining
\be
q_1 = \oint c^1 b_1,\qquad q_2 = \oint c^2 b_2, \qquad q_3 =  \oint c^3  b_3\, .
\ee
Then ${\sf Q}_i$ has degree 1 with respect to $q_i$ and degree zero
with respect to the other gradings.  These three pieces of ${\sf G}^+_0$ mutually
commute and are separately nilpotent, but in general this does not guarantee
that  the cohomology of ${\sf G}^+_0$ is obtained by computing the cohomologies
of the individual $\Q_i$'s successively.
In the case at hand it turns out that we can compute the cohomology (\ref{problem})
as:
\ba  \label{total}
&& H_{{\sf G}^+}(  H_{\Q_F^M + \Q_F^L} (\F_{}  )) \cong H_{{\sf G}^+ + \Q_F^M + \Q_F^L} ( \F_{}  ) \\
 && \cong H_{\Q_1+ \Q_F^M + \Q_F^L}(   H_{\Q_R} ( \F ) ) \cong H_{\Q_1+ \Q_F^M + \Q_F^L}
 (H_{\Q_3} (H_{\Q_2}( \F))) \,.
 \nonumber
\ea
The justification of this claim is  based on a  simple property of double complexes
and can be found in Appendix A.
We see that we can evaluate
the cohomology in the order $\Q_2$, $\Q_3$, $\Q_1 + \Q_F$.
 Here is a schematic outline of the calculation:

{\it $\Q_2$ reduction}:
The BRST operator ${\sf Q}_2$ is associated to a positive root of $\su$,
and computing its cohomology is  similar to Hamiltonian reduction,
except that here we are setting $\beta_L \to 0$ rather
than to a constant.
This step gets rid of the quartet
$\{ b_2,c^2, \beta_L, \gamma_L \}$.

{\it  $\Q_3$ reduction}:
The operator ${\sf Q}_3$ is associated to a ${\frak  \hat u(1)}$  generator of $\su_{p-2} \oplus \su_{-p-2}$.
Taking its cohomology amounts to restricting to the states which
are invariant under this current.
This is similar in spirit the reduction from $SL(2) \times U(1)$ to $SL(2)/U(1)$
described in \cite{Gawedzki:1988hq, Karabali:1989dk}.  This steps essentially kills the second quartet
$\{ b_3, c^3, \beta_M ,\gamma_M  \}$.

We now turn to a more detailed analysis.

\newsubsection{$\Q_2$ reduction}

The operator ${\sf Q}_2 = \oint c^2 \beta_L$ is just the supercharge for the quartet
$\{ b_2 c^2 \beta_L \gamma_L \}$. The standard Kugo-Ojima quartet mechanism applies:
in $\Q_2$ cohomology only the vacuum state survives. Actually
because of the usual phenomenon of picture degeneracy for a $\beta \gamma$ system,
there   are infinitely many vacua, one for each choice of picture.  Vacua in different pictures must be considered physically identical. One can move between different pictures
with the help of
 the picture raising and lowering operators,
\be
Y_L = b_2 \delta(\beta_L) , \qquad Z_L = c^2 \delta(\gamma_L) \, .
\ee
The precise statement  of  $\Q_2$ reduction
is that once we commit ourselves to a choice of picture
 - for example by restricting to the Fock space built on the $SL(2)$ vacuum -
only the vacuum in that picture survives.
In particular for a given picture, $H_{\Q_2}$ is non-trivial for only one value of the grading $q_2$.

\newsubsection{$\Q_3$ reduction}

The operator ${\sf Q}_3$ be viewed as the BRST charge  for the $U(1)$ symmetry
 generated by the current
\begin{equation}
{\scr J} =- \sqrt{ p} \del \varphi + i \sqrt{p} \del x + \gamma_M \beta_M + c^1 b_1 \, ,
\end{equation}
with $c^3, b_3$ the corresponding ghost  and antighost fields.\footnote{We have left out
the  term  $\beta_L \gamma_L - c^2 b_2 = \{ {\sf Q}_2, b_2 \gamma_L \}$
because it acts trivially on $H_{{\sf Q}_2}$.}
The $\Q_3$ cohomology consists of the space of gauge-invariant states.
In particular, since ${\scr J}_0 =\{  \Q_3 ,  (b_3)_0 \}$,
all states in  $H_{{\sf Q}_3}$ must be singlets with respect to the $U(1)$ current.

A familiar way to organize the calculation is to separate
out the ghost zero mode.
To find  $H_{\Q_3}$, we can first
calculate the relative cohomology
$H^R_{\Q_3}$ in the complex annihilated by $(b_3)_0$.
By standard arguments, the relative cohomology is non-trivial only
for degree $q_3 = 0$. Then the absolute cohomology $H_{\Q_3}$
 is simply given by $H_{\Q_3} \cong H^R_{\Q_3} \oplus (c^3)_0 H^R_{\Q_3} $.
It remains to find  a useful representation for $H^R_{\Q_3}$.
We have seen that for  $\alphap = p +1$, the fields
 $\varphi$, $ x$, $ b_1$ and $c^1$ are very reminiscent of the fields of the minimal string and
 we would like to use them as generators of the $H_{{\sf Q}_3}$,
but they are not $U(1)$ invariant. There is a simple remedy
for this - we are going to dress
them up with appropriate powers of
 $\beta_M$ or $\beta_M^{-1}$:
\begin{equation} \label{fields}
\begin{array}{rcll}
e^{-2 \Phi/\sqrt{p}} &=& \beta_M e^{-2 \varphi/\sqrt{p}}  \eol
e^{-2 i X/\sqrt{p}} &=& \beta_M e^{-2 i x/\sqrt{p}}  \eol
B &=& b_1 \beta_M    \eol
C &=& c^1 \beta_M^{-1} \,.
\end{array}
\end{equation}
By a simple counting we see that the fields
$\Phi,  X,  B,  C$ in fact generate all gauge-invariant combinations.
These fields have the remarkable property that
their dimensions are independent of $\alphap$.  The $(B, C)$
fields are fermionic ghosts of dimensions and $(2, -1)$,
and the background charges of $X$ and $\Phi$ are the correct ones
for the matter and Liouville fields of the $(p,1)$ model.
So we seem to have  obtained  simple expressions
for the generators that can be identified exactly with the fields
of the minimal bosonic string.%

This is essentially correct, however
these generators contain some redundancy with respect
to the original degrees of freedom,  introduced in the step of taking the formal
inverse power
 of $\beta_M$.
 To make sense of $\beta_M^{-1}$, we
can bosonize the $\beta_M \gamma_M$-system.  To this end we
 introduce two scalar fields $\rho, \sigma$
with  stress tensor
\begin{equation}\label{}
    {\sf T}_{\rho \sigma} =
    -(\del\rho)^2 + \frac{1-2\lambda}{\sqrt{2}} \del^2\rho - (\del\sigma)^2 +  \frac{i}{\sqrt{2}} \del^2\sigma \, ,
\end{equation}
and write\footnote{The unconventional factors of $\sqrt{2}$ arise from our
choice $\alpha' =1$, which is awkward here but quite useful elsewhere. }
\be
\label{bosoniz}
  \beta_M =  \eta  e^{- \sqrt{2} \rho }\equiv
   e^{ - i \sqrt{2} \sigma- \sqrt{2} \rho}\, ,  \quad \gamma_M = -\del \xi  e^{\sqrt{2} \rho} \equiv
   -i \sqrt{2} \del \sigma e^{i \sqrt{2} \sigma+ \sqrt{2} \rho}\, .
\ee
The $\eta \xi$ fields are fermionic ghosts of dimensions $(1, 0)$. We have quoted
the general formula for  $\Delta(\beta_M ,\gamma_M) = (\lambda, 1-\lambda)$, but in the
following we take $\lambda =\frac{p+1 -  \kappa}{p}$. Now we can set
\be
\beta_M^{-1} = \xi e^{\sqrt{2} \rho}  \,.
\ee
As is familiar, bosonization introduces
one additional zero mode,
the zero mode of $ \xi = e^{i \sqrt{2}  \sigma}$.  Moreover
we encounter the usual picture degeneracy: there are infinitely many  copies of the Hilbert space
of the $\beta_M \gamma_M$ system labeled by the picture $\oint (\beta_M \gamma_M + \eta \xi) = p_\rho- p_\sigma$, where
$p_\rho \equiv  \sqrt{2} \oint \del \rho $,  $p_\sigma \equiv \sqrt{2} \oint i \del \sigma$.
 Since
\begin{equation}\label{}
\{ {\sf Q}_R, b_3-b_2\gamma_L \} =
- \sqrt{ p} \del \varphi +  i \sqrt{ p} \del x + \gamma_M \beta_M + c^1 b_1\, ,
\end{equation}
we can define the picture changing operators
\be
Z_M = e^{ \sqrt{ p} \varphi -  i \sqrt{ p} x - \sqrt{2}\rho -\int c^1 b_1}, \quad
Y_M = e^{ -\sqrt{ p} \varphi +  i \sqrt{ p} x + \sqrt{2}\rho +\int c^1 b_1} \,.
\ee
These operators are $\Q_R$-closed and their derivatives are $\Q_R$-exact.

The fields $\Phi, X, B, C$ are all automatically at zero picture. We still
need to restrict to  the ``small Hilbert space'' of states that do not contain $\xi_0$,
or equivalently to the kernel of $\eta_0$.
It is convenient to use the zero-picture version of $\eta_0$, the nilpotent operator
\begin{eqnarray}
{\sf F} &\equiv &  \oint   \eta \, Z_M
= \oint \beta_M e^{\sqrt{p} \varphi -i \sqrt{p} x -\int c^1 b_1} \eol
 &=& \oint B  e^{\sqrt{p} (\Phi -i  X)} \,.
\end{eqnarray}
We can now state the final result for $H^R_{\Q_3}$:
it consists of  the states generated by  $\Phi, X, B, C$ and in the {kernel}
of ${\sf F}$.\footnote{It is worth noticing
that  ${\sf F}$  is precisely  the fermionic screening operator
$\tilde Q$ encountered in \cite{Bershadsky:1992pe} in the  analysis of
the underlying $N=2$ structure of the minimal bosonic string.
It is also equivalent to the operator usually denoted by $Q_S$ in the context
of topological gravity \cite{Dijkgraaf:1990qw,Li:1990fc}.
We suspect that a deeper
understanding of this structure may involve an $N=4$ topological
algebra \cite{Berkovits:1994vy}, with $\Q_{\rm Vir}$ and ${\sf F}$ as the $N=4$ $G^+$ and $\tilde G^+$
generators.}

\newsubsection{Comparison with the cohomology of the minimal string}

Let us now examine the Felder charges. The
$\su_{p-2}$  screeners are
\be
 \beta_M e^{-2ix/\sqrt{p}} = e^{-2iX/\sqrt{p}} \, ,  \quad
  \beta_M^{-p} e^{2i\sqrt{p}x} = e^{2i\sqrt{p} X} \, ,
\ee
which are just the usual matter screening operators for the $(p,1)$ model  (see (\ref{QFM}) in
 Appendix B).
The $H_3^+$ screeners on the other hand turn out to be exact:
\be
 \beta_L e^{-2\varphi/\sqrt{p}} =
\{ {\sf Q}_R, b_2 e^{-2\varphi/\sqrt{p}} \} \, ,
\ee
and similarly for the $+$ screener.
Thus the $H_3^+$ screener does not reduce to the cosmological constant operator,
which instead descends from
\be
\beta_M e^{-2\varphi/\sqrt{p}} = e^{-2\Phi/\sqrt{p}} \,.
\ee

The last piece $\Q_1$ of the original topological BRST operator $\G^+_0$
will be shown below to be equivalent to
 $ \Q_{\rm Vir}$ of the minimal string up to $\Q_R$
exact terms.
All in all, we can summarize our findings as\footnote{For clarity we omit here
the doubling of the cohomology due to $(c^3)_0$.}
\be \label{finalH}
H_{\G_0^+} (H_3^+ \times S^3) \cong H_{\Q_{\rm Vir} + \Q_F^X}  ( {\rm Ker}_{{\sf F} } ({\cal F}_{X} \otimes
{\cal F}_{\Phi} \otimes{\cal F}_{BC}   ) )\,.
\ee
Here $ \Q_F^X$ denotes the Felder charge for the matter $(p,1)$ model.
We should also recall that the field $X$  inherits from $x$ the restriction to the lattice
of momenta $p_X = \frac{m -1}{2 \sqrt{p} }$, $m \in  \mathbb{Z} $, $m \neq 0$ mod $p$ (see equ.(\ref{lattice})).

We claim that (\ref{finalH})  contains all the expected
``small phase space'' states of the minimal $(p,1)$ string.
As we review in detail in Appendix B, the
cohomology of the $(p,1)$ string\footnote{
We focus in the following on the chiral cohomology
relative to $B_0$, since if that is found
to agree,  the construction of closed string
states (in semi-relative cohomology)  poses no problem. We also restrict to states
obeying the Seiberg bound.}
 consists of infinitely many tachyon states
$T_{r,s}$  and infinitely many ground ring
states $G_{r,s}$,  with $1 \leq r \leq p-1$, $s \geq 1$. By definition, the small
phase space is spanned by the states with $s =1$.
 Using the explicit expressions for $T_{r,1}$ and $G_{r,1}$,
it is straightforward to check
that  these states are elements of  (\ref{finalH}): they
are annihilated by $\sf F$ and define non-trivial
representatives of the cohomology classes.
The restriction of the complex to the kernel of ${\sf F}$ is crucial. Without this restriction
the cohomology (\ref{finalH}) would be empty, since   for  $(p,1)$ matter
the Felder complex is exact. This implies
that for example   $T_{r,1} =(\Q_{\rm Vir} + \Q_F^X ) (\Lambda_{r,s})$ for some state
$\Lambda_{r,1}$;  but one finds $\Lambda_{r,1} \not \in {\rm Ker}_{\sf F}$.

To obtain the ``gravitational descendants'' of the minimal string,
which are the states with $s > 1$,
we will need to enlarge the spectrum of the theory on $H_3^+ \times S^3$
by allowing for ``spectral flowed'' states (long strings).  Including
such states is actually a necessity for  consistency
of string theory on  $H_3^+ \times S^3$ \cite{Maldacena:2000hw}.
The relation betwen gravitational descendants and long strings will be discussed
in section 4.4.  Ultimately we will find that all the states are correctly matched.

\newsubsection{Similarity transformation}

 We will now establish a correspondence
 between the twisted $N=2$ structure of the topological string  and
 the underlying twisted $N=2$ structure of the minimal bosonic string.
 It will follow from this map that the
rules for computing $N=2$ amplitudes on $H^+_3 \times S^3$ coincide with
the usual rules for minimal bosonic string amplitudes, up to some additional
insertions of picture changing operators.

The strategy will be to perform a similarity transformation on the $N=2$ algebra given
in (\ref{N2explalpha}), repeated here for reference:
\begin{eqnarray}
\sf{T} &=&
 {1\over 2p}(2 j_3 j_3 + j_+ j_- + j_- j_+ - 2 k_3 k_3 - k_- k_+ - k_+ k_-) \eol
& &+ {1\over 2}\left(\del c^i b_i - c^i \del b_i \right)
+ {\alphap \over 2p}\del J_3^{\rm tot}
\eol
\sf{G}^+ &=&  c^1(j_-) + c^2( k_+) + c^3( j_3 +  k_3 - c^2 b_2 + c^1 b_1)
    \eol
\sf{G}^- &=& {1\over p}\left(
b_1 j_+ -b_2  k_- + b_3(j_3 - k_3 + c^2 b_2 + c^1 b_1)
 + \alphap \del b_3 \right)  \eol
\sf{J} &=& c^i b_i  + {\alphap-2 \over p} J_3^{\rm tot} \,.
\end{eqnarray}
We take as generator of the similarity transformation the operator
\begin{equation}\label{}
{\sf L} =  -\oint (c^1 j_+^{-1})\left( -b_2  k_- + b_3( j_3 -
k_3 + c^2 b_2 + c^1 b_1) + \alphap \del b_3 \right) \,.
\end{equation}
Then we find\footnote{
We have checked this expression up to derivatives in $\beta_M$ (which arise through normal ordering)
and passed it through some consistency checks.}
\begin{equation}\label{}
\exp {\rm Ad}({\sf L}) {\sf G^+_0} = {\sf Q}_R + p\, {\sf Q}_{\rm Vir} + \{ {\sf Q}_R,{\bullet} \}
\end{equation}
where ${\sf Q}_{\rm Vir}$ is just the usual Virasoro BRST operator
of the minimal string theory,
\be
{\sf Q}_{\rm Vir} = \oint C(T_\Phi + T_X + \half T_{BC})\, .
\ee
Next we consider ${\sf G}^-$. This generator plays the analogue of the $B$ ghost
in bosonic string theory. A short calculation shows that it is in fact
exactly equivalent to $B$,
\begin{eqnarray}\label{}
\exp {\rm Ad}({\sf L}) {\sf G^-} &=& \frac{1}{ p}(b_1 j_+  ) = \frac{1}{ p} B \,.
\end{eqnarray}
This has the important consequence that
 the integration over the moduli space of Riemann surfaces
and the definition of semi-relative cohomology are
the same both in the topological string and in the
minimal string. Of course, applying the similarity transformation
to the commutator of ${\sf G}^+_0$ and ${\sf G}^-$, we find the expected stress tensor
of the minimal string up to ${\sf Q}_R$-exact terms.

Finally  the $U(1)_R$ current is found to be invariant,
\be
\exp {\rm Ad}({\sf L}) {\sf J} = {\sf J} \,.
\ee
The current $ {\sf J}$
differs from the ghost number current $ C B$ of the minimal string.
This is actually expected,
since the  $U(1)_R$ anomaly prescribes
how we should saturate the background charges.
 Even though two $\{ b,c,\beta, \gamma\}$ quartets
effectively decouple, we still must insert
picture changing operators to saturate their fermionic zero modes.
Taking $\kappa \equiv p+1$,
at genus $g$, we need
 $(g-1)$    insertions of $Z_L \bar Z_L$  operators and
$(g-1)$  insertions of $Z_3 \bar Z_3 \equiv \int d^2 z\,  b_3 \, \bar b_3$.

\newsubsection{General $(p,q)$ models?}

At a formal level,
the manipulations of this section work equally well
with the replacement $p \to p/q$.
Then the $\Q_R$ reduction yields precisely the field
content of  the $(p,q)$ minimal model coupled to gravity.
All the expected $(r,s)$ matter representations of the minimal
string arise provided  start in the ``upstairs'' theory with a direct
sum of $\F_{m, n}$ Fock spaces with $m \in \mathbb{Z}$, $m \neq 0$ mod $p$,
 $1 \leq n \leq q$ (see equ. \ref{lattice}).
So the question arises whether
 we can also give minimal $(p,q)$ strings an interpretation as a
string theory on a Euclidean space with $H$-flux.
For $H_3^+$ the $H$-flux need not be quantized,
so fractional level already makes sense. Of course
the issue is how to interpret  $SU(2)$ at fractional level.\footnote{
As we are going to see shortly, there is a sense in which
the topological theory discussed so far computes
physical amplitudes in the sector of one long string.
It is tempting to speculate that the $(p,q)$ models should
correspond to the sector of $q$ long strings. This suggestion is due
to N.~Seiberg.}

\newsection{What does the topological string compute?}

In this section we  turn our attention to IIB string theory on
$H_3^+ \times S^3 \times M_4$. In  close analogy with the Calabi-Yau case,
 the topological string theory on $H_3^+
\times S^3$ computes a set of special
amplitudes of spacetime chiral primaries in the IIB
theory, in the presence of background RR 5-form flux. The
additional RR insertions are responsible for twisting the
worldsheet theory.
These special amplitudes are at string
tree-level from the viewpoint of the physical theory. The
perturbative expansion of the topological theory corresponds to an
expansion in powers of the 5-form RR flux. A difference
with respect to the Calabi-Yau case is the extra twist
by $\partial J_3^{\rm tot}$.  This gives additional insertions,
which turn out to be insertions of ``long string'' vertex operators.
The topological theory computes amplitudes in the
sector of one long string.

In the remainder of this section we work out the
correspondence between spacetime chiral primaries on $H_3^+ \times
S^3 $ and the physical operators in the minimal string
theory, both for the tachyons and the ground
ring elements.
Finally we discuss a family of exact deformations
of the  $H_3^+ \times
S^3 $ sigma model.

\newsubsection{Amplitudes}

We would like to understand exactly which physical amplitudes can
be reduced to the topological string. The analysis is almost
identical to \cite{Bershadsky:1993cx} and \cite{Antoniadis:1993ze}
and so we will be  brief. We can focus on the partition function,
since the addition of external operators is straightforward. The
physical partition function is given by the path-integral over the
full set of ten-dimensional fields: the matter fields of the
supersymmetric sigma model on $H_3^+ \times S^3 \times M_4$ plus
the diffeomorphism ghosts and superghosts $\{ b,  c,  \beta,  \gamma \}$.
The topological partition function is given instead by the
path-integral over the six dimensional fields only, with the
fermions having twisted dimensions. Finally the measure on the
moduli space of Riemann surfaces is different in the two theories.
Nevertheless, in analogy with the Calabi-Yau case, one can show
that the topological partition function equals the physical
partition function at the only cost of introducing some extra
 insertions.  Because of the relation (\ref{TT}) between
 the physical and the twisted stress tensors, we need
 to make insertions of two kinds: ``graviphoton'' vertex
 operators implementing the usual twist by $\frac{1}{2} \partial (c^i b_i)$;
 ``spectral flow'' operators implementing the twist by $\partial J_3^{\rm tot}$.

 The operators analogous to the graviphotons of the Calabi-Yau case  are
\be T^\pm = e^{-\phi/2-\tilde{\phi}/2}e^{\half \int c^i b_i+ \half
\int \tilde{c}^i \tilde{b}_i}
 \Sigma^\pm \bar{\Sigma}^\pm \,.
\ee
Here $\phi$ is the boson arising from the usual bosonization of
the $\beta \gamma$ superghosts \cite{Friedan:1985ge}. The operators  $\Sigma^\pm$ are
the spectral flow operators for the fermions on $M_4$. In the case
of $M_4 = T^4$,  denoting the four free fermions by $\psi_i$,  we
can write \be \Sigma^\pm = e^{ \pm \frac{1}{2}\int \psi_1 \psi_2
+ \psi_3 \psi_4 } \,. \ee One can can check that $c  \bar cT^\pm$
 are physical  vertex operators for the superstring.
  Indeed we have $\Delta(T^+)=1$ and $T^\pm$ has no
$z^{-3/2}$ pole in its OPE with the $N=1$ worldsheet supercurrent
${\sf G}$.  Just as in the Calabi-Yau case, $T^\pm$   correspond
to the 5-form RR flux,
with three indices along  the ``internal'' six dimensional manifold
$H_3^+ \times S^3$, and two indices along the $M_4$. They
 are also closely related to {\it spacetime} supercurrents.
For $H_3^+ \times S^3$, the left-moving spacetime supercurrents
are \cite{Giveon:1998ns} \be {\cal S}_{\vec \epsilon} =
e^{-\phi/2}e^{\half \int  \sum_i^3 \epsilon_i c^i b_i + \epsilon_4
\psi_1 \psi_2 + \epsilon_5 \psi_3 \psi_4} \, , \ee
 where $\epsilon_i = \pm 1$ with the constraints $\prod_{i=1}^3 \epsilon_i =
 \prod_{i=1}^5 \epsilon_i =1$. So we see that $T^\pm$ are products
 of a left-moving and a right-moving
{spacetime} supercurrent .

The additional  ``spectral flow'' operator is defined as
\be
U \equiv \exp( - \int J_3^{\rm tot}) \, ,
\ee
and its interpretation will be discussed shortly.
With this notation in place, we can formulate
the basic claim:
 the genus $g$ topological partition
function $F_g$
 corresponds to the physical amplitude
\be
F_g = \int_{{\cal M}_g} \vev{  \Theta_g \,
R^2 \, \prod^{g-1} \int \! d^2 z \; T^+
\prod^{g-1}\int  \! d^2 z \;T^- \,
 \prod^{3g-3}| \,\delta(\beta){\sf G}|^2\;  \prod^{3g-3}| b(\mu_i)|^2 }_g^{\rm
physical} \, ,
\ee
with
\be
\Theta_g \equiv (U \bar U)^{2 - 2g} \, (Z_L \bar Z_L)^{g-1}
(Z_3 \bar Z_3)^{g-1} \,.
\ee
As in \cite{Bershadsky:1993cx}, $R^2$ denotes  the extra insertion
of
 a $\psi^8$ expression  needed to account properly for the fermion
 zero modes. The proof of this claim imitates the arguments of \cite{Bershadsky:1993cx} and \cite{Antoniadis:1993ze}. Here is a very brief sketch:  The $T^\pm$ insertions
 are responsible for twisting the fermions and changing the background
 charge of the $\beta \gamma$ system, in such a way that the
 path-integral over the $M_4$ fields and the $bc \beta \gamma$ system
 can be trivially performed.  Similarly, the insertions of $U $ implement
 the twist by $\partial J_3^{\rm tot}$. The insertions of $Z_L$ and $Z_3$
 are needed to saturate fermion zero modes of $c^2 b_2$ and
 $c^3 b_3$, as discussed in  section 3.7.
  Writing
${\sf G} = {\sf G}^+ + {\sf G}^-_{\alphap=0} + {\sf G}_{M_4}$, we
see that only ${\sf G}^-$ can contribute due to the anomalous
background charge for ${\sf J}$. So finally we have
\be F_g = \int_{{\cal M}_g}\vev{ \prod^{3g-3}| {\sf
G}^-_{\alphap=0}(\mu_i)|^2\, (Z_L \bar Z_L)^{g-1}
(Z_3 \bar Z_3)^{g-1}}^{\rm topological}_g \,.
\ee
Apparently there is a discrepancy between
the value of $\kappa= p+1$ in the twisted
stress tensor, and the value of $\kappa = 0$ in ${\sf G}^-$.
This is not a problem: the
saturation of fermionic
zero modes is such that we can replace
${\sf G}^-_{\kappa = 0}$ with ${\sf G}^-_{\kappa = p+1}$
because
 the $\kappa$-dependent term
 does not contribute. Equivalently,
under the similarity transformation,
 ${\sf G}^-_{\kappa = 0}$
gets  mapped to $(B - (p+1)\partial  b_3)/p$,
and again the term proportional to $\partial b_3$ does not contribute.

Finally we turn to the interpretation of the
operator $U$. This is exactly the supersymmetric spectral
flow operator constructed in \cite{Argurio:2000tb} following \cite{Maldacena:2000hw}.
The intuitive picture
is that an insertion of $U^w$ creates a ``long string'' - a worldsheet that wraps
$w$ times the boundary of $H_3^+$; it also adds
angular momentum around the $S^3$, in such a way that spacetime
supersymmetry  is preserved.

\newsubsection{Vertex operators for $H_3^+ \times S^3$}

Our task is now to identify the physical states of the topological
theory, defined as cohomology classes of ${\sf G}^+_0$, with
physical states of the ten-dimensional theory, defined as
cohomology classes of the usual superstring BRST charge
$\Q_{10d}$.  To this
end, we review in this subsection some general properties of
vertex operators in  $H_3^+ \times S^3$
\cite{Giveon:1998ns,Kutasov:1999xu}, emphasizing their relation
with bulk-to-boundary propagators. In the following subsection we
will specialize to vertex operators for spacetime chiral
primaries \cite{Kutasov:1998zh,Argurio:2000tb},
which are the natural candidates to match with the
topological observables. This subsection and the next are almost entirely
review of well-known material.

The WZW model on $H_3^+$ with a pure imaginary $B$-field is
described by the following action
\be S =  \int d^2 z\, \left( \del \varphi \delb \varphi +
e^{2\varphi}\delb \gamma \del \bar{\gamma}  \right)\, . \ee
The coordinates $\gamma, {\bar \gamma}$ and $\varphi$ satisfy
$\gamma^* ={\bar \gamma}, \varphi^* = \varphi$. This action is
real and positive definite on a Euclidean worldsheet. The zero
modes of the currents are given by the left- and right-invariant
vector fields
\ba \oint k_+ = -\del_\gamma, \qquad \oint k_3 = -\gamma
\del_\gamma + \half \del_\phi, \qquad \oint k_- = \gamma^2
\del_\gamma - \gamma \del_\phi - e^{-2\phi}\del_{\bar{\gamma}}
\eol \oint {\bar{k}}_+ = -\del_{\bar{\gamma}}, \qquad \oint {\bar
k}_3 = -{\bar \gamma} \del_{\bar \gamma} + \half \del_\phi, \qquad
\oint \bar{k}_- = {\bar \gamma}^2 \del_\gamma - {\bar \gamma}
\del_\phi - e^{-2\phi}\del_{{\gamma}} \,. \ea
Here we have put a bar on the currents to denote right-movers on the worldsheet. Let
us also define the total currents
\begin{equation}
\begin{array}{ll} \label{currents}
{\sf K}_3 = k_3 - c^2 b_2    &\qquad {\sf J}_3 = j_3 + c^1 b_1
\eol {\sf K}_+ = k_+ + (c^3- {1\over p}b_3) b_2  &\qquad {\sf J}_+
= j_+ - c^1(p c^3 + b_3)  \eol {\sf K}_- = k_- + (pc^3 - b_3)c^2
& \qquad {\sf J}_- = j_- - b_1(c^3 + {1\over p}b_3)
\end{array}
\end{equation}
They form an $\su$ current algebra of level $-p$ and level $p$
respectively. The zero modes of the currents (\ref{currents}) are
conserved charges in the spacetime theory. It is convenient to use
these charges to organize the spectrum of string theory on $H_3^+
\times S^3$. In particular, according to Brown and Henneaux \cite{Brown:1986nw}
the spacetime theory is itself a conformal theory and comes with its
own set of Virasoro generators $\L_n$ which are distinct form the
worldsheet Virasoro generators. As  described in \cite{Giveon:1998ns},
 the zero modes of ${\sf K}_+, {\sf K}_-,{\sf K}_3$ map to the
spacetime Virasoro generators, and the zero mode of ${\sf J}_3$ to
the space time R-charge. The spacetime Virasoro algebra is given
by
\be [ \L_n, \L_m ] = (n-m) \L_{n+m} + {c^{st} \over 12} (n^3 - n)
\delta_{n+m} \,  , \quad c^{st}= 6 Q_1 Q_5 \,. \ee
In order for this to agree with our commutation relations for
${\sf K}_+, {\sf K}_-,{\sf K}_3$, we need either $\{ \L_1,\L_{-1},\L_0 \} \to \{
\oint {\sf K}_+, -\oint {\sf K}_-, -\oint {\sf K}_3 \}$ or to $\{
-\oint {\sf K}_-, \oint {\sf K}_+, \oint {\sf K}_3 \}$. Since we
wish to identify $\L_{-1} $ with the generator $\del_\gamma$ of
translations on the boundary of $H_3^+$,
we pick the second possibility,
\be \{ \L_1,\L_{-1},\L_0 \} \to \{ -\oint {\sf K}_-, \oint {\sf
K}_+, \oint {\sf K}_3 \}\, . \ee
In particular, the (left-moving) spacetime weight is measured by
$\L_0 = +\oint {\sf K}_3$. We will denote the eigenvalues of
$\L_0$ by $h$. Similarly the spacetime R-charge is measured by \be
q \equiv -\oint {\sf J}_3 \,. \ee

The vertex operators for the physical string states are built from
the free fermions and primaries of the two ${\frak su}(2)$ current
algebras (and 10d ghosts and superghosts $b, c, \beta,\gamma$).
The primaries $\Phi_{jm}(z)$ for $H_3^+$ fall in certain
representations of  ${\frak su}(2)$ at negative level. The
relevant ones for us are the principal discrete representations:
\ba {\cal D}_j^- & : & \quad m = -j, -j-1, \ldots  \eol {\cal
D}_j^+  & : & \quad m = j, j+1, \ldots \ea
These representations are labeled by the worldsheet dimension (which coincides
with the Casimir)
\be
 \Delta = - {j(j-1)\over p} \, ,
\ee
and states are labeled by their eigenvalue $m$ under $k_3$.

To discuss strings on $H_3^+$ it is convenient to exchange the
label $m$ for the isotopic spin coordinate $x$. We can interpret
$x$, $\bar x$ as parametrizing the boundary of $H_3^+$. The
vertex operator $\Phi_j(z|x)$ is then required to satisfy the OPEs
\ba k_a(z) \Phi_j(w,\bar{w}|x,\bar{x}) &\sim& {D_a \over z-w}
\Phi_j(w,\bar{w}|x,\bar{x}) \ea
with
\be D_+ = {\del\over \del x}, \qquad D_3 = x{\del\over\del x}+j,
\qquad D_- = x^2 {\del\over \del x} + 2 j x\,. \ee
This implies the following expression:
\be \Phi_j(x) = {1-2j \over \pi}\left({1\over |\gamma - x|^2
e^\varphi + e^{-\varphi}}\right)^{2j} \,. \ee
This is exactly  the standard bulk-boundary propagator written in
Poincar\'{e} coordinates,
for a field of spacetime dimension $h=j$
inserted at $(x, \bar x)$ on the boundary. It admits the following
expansion:
\be \Phi_j(x) = e^{2(j-1)\varphi}\delta^2(\gamma-x) + \ldots +
{e^{-2j\varphi}\over |\gamma - x|^{4j} } + \ldots \ee

Now we turn to a free field description. We introduce two new
variables $\beta_L, \bar{\beta}_L$ and
 rewrite the sigma model action as%
\be S = \int d^2 z \left( \del \varphi \delb \varphi + \beta_L
\delb \gamma_L + \bar{\beta}_L \del \bar{\gamma}_L - e^{-2\varphi}
\beta_L \bar{\beta}_L \right) \,. \ee
By classically integrating out $\beta_L$ and $\bar{\beta}_L$  we
recover the original action. Taking into account the quantum
measure introduces a linear dilaton and renormalizes some of the
terms. The correct confomally invariant action is
\be S =  \int d^2 z \left( \del \varphi \delb \varphi - {2\over
\sqrt{p}} R \varphi + \beta_L \delb \gamma_L + \bar{\beta}_L \del
\bar{\gamma}_L - e^{-\frac{2\varphi}{\sqrt{p}}} \beta_L
\bar{\beta}_L\right) \, . \ee
Now we can treat $ \beta_L \bar{\beta}_Le^{-2 \varphi/\sqrt{p}}$
as a screening operator and use $\varphi,\beta_L,\gamma_L$ as free
fields. Writing out the currents yields the Wakimoto
representation used in section 3. Calculations with this free field representation
are valid when the correlation functions are supported near the boundary
$\varphi \to \infty$, and merely use the leading term in the bulk-boundary propagator:
\be\label{lowestweight}
  \Phi_j(0) \to V^{H_3^+}_{j,j} \equiv
\delta(\gamma_L) e^{2(j-1)\varphi/\sqrt{p}} . \ee
This operator is annihilated by $\L_{n \geq 1 }$, and  it is
therefore a spacetime Virasoro primary inserted at $x=0$. Acting on
it with $k_+$ maps out a ${\cal D}_j^+$ representation.

There is another set of vertex operators which use the ${\cal D}_j^-$
representation:
\be
  V^{H_3^+}_{j,-j} \equiv e^{-2j\varphi/\sqrt{p}} \,.
\ee
These operators have regular OPE with $k_+$, therefore they are
annihilated by $\L_{ -1}$, in fact by all
$\L_{n\leq -1}$.
Acting on $V^{H_3^+}_{j,-j}$ with $k_-$ maps out the ${\cal D}_j^-$
representation. These operators correspond to normalizable modes in spacetime -
they carry the ``$\Delta_-$ dressing''
in the usual AdS/CFT language.
Their insertion in correlation functions should be interpreted
as infinitesimal changes of the {\it state} of the  theory,
specified by the vacuum expectation values of the spacetime operators.

Finally we need to consider Wakimoto vertex operators for $S^3$.
The primary vertex operators for the spin $j$ representation of
$\su_{p-2}$ are given by
\be V^{S^3}_{j,m} = \gamma_M^{j-m} e^{2ijx/\sqrt{p}} \, , \qquad  0
\leq j \leq p/2 -1 \, , \quad \ -j \leq m \leq j\, , \label{VS}
\ee
with worldsheet dimensions
\be \Delta(V^{S^3}_{j,m}) = {j(j+1)\over p} \,. \ee
They satisfy the following OPEs:
\ba j_+(z) \cdot V^{S^3}_{j,m}(w) &=& {-j+m \over z-w}
V^{S^3}_{j,m+1}(w)  \eol j_3(z) \cdot V^{S^3}_{j,m}(w) &=& {2 m
\over z-w} V^{S^3}_{j,m}(w)   \eol j_-(z) \cdot V^{S^3}_{j,m}(w)
&=& {-j - m \over z-w} V^{S^3}_{j,m-1}(w) \ea
In particular, for a highest weight state we have the simple
expression
\be V^{S^3}_{j,j} =  e^{2ijx/\sqrt{p}}\,. \ee
For lowest weight states we could take $m =-j$ in (\ref{VS}), but
it turns out that a different (equivalent) representation makes
contact more directly with the minimal string. By analogy with
(\ref{lowestweight}) we represent lowest weight states for $S^3$
as
\be \tilde{V}^{S^3}_{j,-j} \equiv \delta(\gamma_M)
e^{-2i(j+1)x/\sqrt{p}} \,. \ee

\newsubsection{Chiral primaries}

The spacetime theory has $(4, 4)$ supersymmetry. Here we are
defining  chiral operators with respect to a given subgroup
$U(1)_R  \subset SU(2)_R$ of the R-symmetry. Each short multiplet
of $N=4$ contains precisely one chiral and one anti-chiral primary
with respect  to this $U(1)_R$.
 In defining the topological theory, we picked
a preferred $U(1)_R$ (generated by ${\sf J}_3$) in the choice of
the complex structure.

Since ${\sf J}_3^{\rm tot} = {\sf K}_3 + {\sf J}_3$ is
$\G_0^+$-exact, a cohomology element of the topological string
necessarily obeys $h = q$. There are two possibilities. If $h = q
> 0$, the vertex operator is a (spacetime) chiral primary inserted
at $x=0$; it is the lowest weight state of a ${\cal D}_j^+$
representation for $H_3^+$, and the lowest weight state for a
finite representation of $S^3$. If $h=q < 0$,
it is the highest weight state of a ${\cal D}_j^-$ representation for $H_3^+$, and the highest
weight state for a finite representation of $S^3$; it corresponds
to turning on a vev.

Vertex operators for spacetime chiral primaries have been
classified, so we can check the known list of such primaries for
candidates that can descend to the minimal string. For the purpose
of the topological string we are interested in operators in the
Neveu-Schwarz sector. Then one finds the following two series
\cite{Kutasov:1998zh,Argurio:2000tb}\footnote{ In the cited papers
they are denoted by ${\cal W}_j^-,\ {\cal X}_j^+$. The
superscripts have no meaning for us, so we will leave them out.}
\ba NS: & {\cal W}_j,\ {\cal X}_j \ea
These are the left-movers only. The operator ${\cal W}_j$ has an
index along $H_3^+$, and ${\cal X}_j$ has an index along $S^3$. By
combining left- and right-movers, one obtains various modes of
$g_{\mu\nu} + b_{\mu\nu}$ on $H_3^+ \times S^3$. The index $j$
comes in half-integer steps and labels the spacetime dimension, or
equivalently the R-charge, as follows:\footnote{We write only
the matter part of the operators. The full  operators are obtained
by adding the usual ghost and superghost factor $c e^{-\phi} $.}
\ba {\cal W}_j = c^2\,  V^{H_3^+}_{j+1,j+1} \,V^{S^3}_{j,-j}: & \
h =q =  j, & \quad l = 0, \ldots, {p-2 \over 2} \eol {\cal X}_j =
b_1 \, V^{H_3^+}_{j+1,j+1}\, V^{S^3}_{j,-j}: & \ h = q = j+1, &
\quad l = 0, \ldots, {p-2 \over 2}\,. \ea
We are also interested in the operators with
$h = q < 0$. These are given by
\ba {\widetilde {\cal W}}_j = b_2 \, V^{H^+_3}_{j+1,-j-1}
V^{S^3}_{j,j}: & \ h = q = -j, & \quad l = 0, \ldots, {p-2 \over
2} \eol {\widetilde {\cal X}}_j = c^1 \, V^{H_3^+}_{j+1,-j-1}
V^{S^3}_{j,j}: & \ h = q=-j-1, & \quad l = 0, \ldots, {p-2 \over
2} \,. \ea
This is however not the full story.
It has been demonstrated in
\cite{Maldacena:2000hw} that apart from the usual vertex operators
based on ${\cal D}_j^-$ and ${\cal D}_j^+$, string theory on
$H_3^+$ has additional physical vertex operators that can be
obtained by spectral flow. Spectral flowed vertex operators
correspond to ``long strings'' wrapping the boundary of $H_3^+$ multiple times.
The resulting
``winding number'' $w$ need not be conserved in correlation
functions because $H_3^+$ is a contractible space.
 In the context of superstring theory on $H_3^+ \times S^3$ it is natural
 to consider  spectral flow operators
which are local with respect to the spacetime supercharges.
The operator relevant for us is
\be U = \exp(-\int { J}^{\rm tot}_3) \, . \ee
This operator increases $h$ and $q$ by $\delta h =\delta q= p/2$.
By repeated application of $U$, one finds finds the
following towers of chiral primaries:
\ba {\cal W}^w_l &  \quad  l = 0, \ldots, {p-2 \over 2},\ w \geq 0
\eol {\cal X}^w_l &  \quad l = 0, \ldots, {p-2 \over 2}, \ w\geq
0. \ea
We have restricted ourselves to $w \geq 0$. Applying spectral flow
in the opposite direction yields ${\cal D}_j^{-, w}$ , because ${\cal D}_j^-$ and ${\cal D}_j^+$ are
related by one unit of spectral flow:
\be {\cal D}_j^{+,w=-1} = {\cal D}_{p/2 - j}^-  \, . \ee

\newsubsection{Matching with the minimal string}

We have discussed a
list of vertex operators that can potentially survive in the topological
theory. Since these are ghost number one states
(with respect to the $c b$ ghosts of the 10d string theory),
they should correspond to ghost number one
states of the minimal string (with respect to the $C B$ ghosts
of the minimal string).  This is necessary if
the counting of  Riemann surface moduli in correlation functions is
to work out correctly.

The spectrum of the minimal string is
discussed in detail in Appendix B.
At ghost number one (in the relative, chiral cohomology)
one has the
$p-1$ tachyons and their gravitational descendants,
\be T_{n= 2j + w+1}  =\hat y^w T_{2j+1}, \qquad j=0, \ldots, {p-2\over 2},\ w\geq
0 \, . \ee
The  $T_1$ tachyon is also known as the puncture operator, and the
$T_{p-1}$ tachyon is also known as the cosmological constant
operator. As we have mentioned, one needs to consider both chiral
primaries with $h >0$ and $h <0$. Let us start with ${\widetilde
{\cal X}}_j$. In  the Wakimoto representation it can be written as
\ba {\widetilde {\cal X}}_j
  &=&
 c^1 \, V^{H_3^+}_{j+1,-j-1}\,  V^{S^3}_{j,j} \eol
 &=& c^1 e^{-2
j (\varphi - i x)/\sqrt{p} - 2 \varphi/\sqrt{p}} \eol
 &=& Ce^{-2 j
(\Phi - i X)/\sqrt{p} - 2 \Phi/\sqrt{p}} \ea
Remarkably these are precisely the expressions for the tachyons
$T_{p-1-2j}$ of the minimal string theory. As we discussed, these
operators are interpreted as normalizable modes in $H_3^+$. They
are of course non-normalizable in Liouville theory. This
difference arises because of the shift in the background charge of
$\varphi$ due to the $\partial J_3^{\rm tot}$ twist.

One can also check that $\widetilde {\cal W}_j$ and ${\cal X}_j$
are {\it not} in the cohomology of ${\sf G}_0^+$.

\begin{figure}[t]
\begin{center}
\leavevmode\hbox{\epsfxsize=5.5in
\epsffile{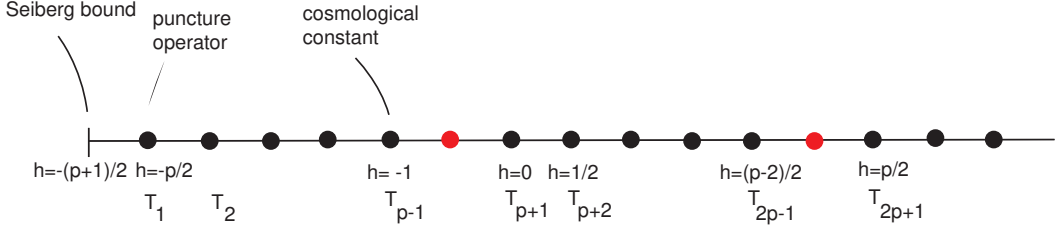}}\\[3mm]
\caption{ \it The spectrum is organized according to increasing
Liouville momentum, or equivalently increasing spacetime
dimension. The red dots indicate the missing states.
\label{dots}}
\end{center}
\end{figure}
Next we consider ${\cal W}_j$. We can write them as
\ba {\cal W}_j &=& c^2 \, V^{H_3^+}_{j+1,j+1}\, V^{S^3}_{j,-j}
\eol
 &=&
c^2 \delta(\gamma_L) \delta(\gamma_M) e^{2 (j+1) (\varphi - i
x)/\sqrt{p}}e^{-2  \varphi/\sqrt{p}} \ea
Comparing the Liouville momenta of these operators with the
spectrum for the minimal string suggests that the ${\cal W}_j$
should be identified with the first group of descendants $\hat y
T_{2j+1}$.

To understand how this comes about, let us first look more closely
at the spectral flow operator \cite{Argurio:2000tb}:
\be U(z) = \exp(-\int^z { J}_3^{\rm tot}) = c^2 b_1
\delta(\gamma_L)\delta(\gamma_M)\exp( \sqrt{p}( \varphi - i
x)) \, .\ee
It is precisely the picture changing operator $Z_M
Z_L$ that we have encountered before. Essentially
$U$ plays the role of the ground ring generator
$\hat y$ - it has the correct $x$ and $\varphi$
dependence -  but being a picture
changing operator, formally it does not create any new states . A similar situation was
encountered in \cite{Mukhi:1993zb}. The resolution
is to consider the closely related operator
\be \tilde{U} = U e^{-i\sqrt{2}\sigma} = c^2
\delta(\gamma_L)B\exp( \sqrt{p}( \Phi - i  X)) \, ,\ee
where we replaced $\delta(\gamma_M) \to \beta_M$  by combining
with $e^{-i \sqrt{2}\sigma}$. This operator is still not quite
the one we need since while
its total ghost number is zero, it has $q_1 = -1$ and $q_2 = +1$.
We should then apply a descent procedure
to obtain an operator with $q_1 = q_2 = 0$.
It is not hard to check that
\be {\{} {\sf Q}_{\rm Vir},\tilde{U} {\}} = \{ {\sf Q}_R,
\theta(\gamma_L) \hat y \} -  \del( c^2 \delta(\gamma_L)
\hat y) \, .\ee
Therefore  $\tilde{U}$ is responsible for creating the descendants.

Now it is easy to check that
\be U  \widetilde {\cal X}_j = {\cal W}_j \ee
and so by dressing up $\delta(\gamma_M) \to \beta_M$ and applying the descent procedure,
the operators ${\cal
W}_j$ should indeed be regarded as the first descendants of
the tachyons. Further acting with $\tilde{U}$ yields the remaining
descendants. These correspond to honest long string states in
$H_3^+$ which cannot be seen in supergravity.

Figure \ref{dots} summarizes the correspondence between
the ghost number one states of the minimal string and the
chiral primaries in $H_3^+ \times S^3$. A striking feature is the presence of
gaps in the spectrum: every $p$ steps, a state is missing.
This is very natural from the viewpoint of the  $p$-KdV integrable
hierarchy, where every $p$-th flow parameter is redundant.
We see that this emerges naturally in the reduction from $H_3^+ \times S^3$.
 Physically, the absence
of these states has been a bit of a mystery for string theory on $H_3^+ \times S^3$.
The holographic CFT on the boundary of $H_3^+$, a deformation
of the symmetric product ${\rm Sym}^{Q_1 Q_5}(M_4)$
 superficially appears to contain such states. It has been suggested
that their absence may be related to the singular behavior of this
CFT \cite{Seiberg:1999xz}. Heuristically, at the point in moduli
space at which we are working, there is no cost in energy for the
system to emit a long string. This leads to a continuum of states
above a certain threshold and it has been proposed that the
missing states may be related to  this continuum. Does the
correspondence with the minimal string give any insight into this
issue? In the minimal string, some of the gaps in the closed
string spectrum are believed to be filled by open string states
\cite{Martinec:1991ht}. For example, the first missing state has
precisely the Liouville dressing to be identified with the
operator that couples to the {\it boundary} length - the boundary
cosmological constant that can be turned on on
 an FZZT brane  \cite{Martinec:1991ht}. An analogous interpretation becomes
 viable for string theory in $H_3^+ \times S^3$:
the first missing state has the correct $\varphi$ dressing to be
identified with the boundary screener of an $H_2^+$ brane.

\newsubsection{Ground ring}

Besides the tachyons, which carry ghost  number one,
the minimal string has physical states at ghost number zero,
the ground ring elements
$G_{n= r + ps } = \hat x^{r-1} \hat y^{s-1}$,
$1 \leq r \leq p-1$, $s \geq 1$. These states are clearly
in the $\G_0^+$ cohomology of the topological
theory. The question is what is their intepretation
in the physical theory. It is natural to expect that they get lifted to ground ring
elements of the IIB theory, that is  cohomology classes of the
 BRST operator $\Q_{10d}$ carrying zero ghost number
 with respect to 10d $c b$ ghosts.  The explicit expressions
 of $\hat x$ and $\hat y$ fail to be annihilated by $\Q_{10d}$,
 but it is possible to add  improvement terms involving
 the 10d $b c \beta \gamma$ ghosts such that they
 become elements of the $\Q_{10d}$ cohomology.
One could have anticipated the existence of a ground ring structure in $H_3^+ \times S^3$ just
 from the representation theory of the $\su_{p-2}$ current algebra:
 a ground ring element must exist for each primitive null over a primary,
 by  a generalization to current algebras of the
 mechanism \cite{Lian:1991gk,Bouwknegt:1991yg,Govindarajan:1992kv} reviewed in Appendix B.

 It is clear that the ground ring states  can be constructed
 by a descent procedure entirely analogous to (\ref{descent}),
 \be
 \Q_{10d}\, | {\cal G}_n  \rangle =    c e^{-\phi}\, \Q^M_F  |T_n \rangle  \,.
 \ee
 The operators $ {\cal G}_n$ are the ``improved'' versions
 of $G_n$ that we are after.
 They generate a $W_p$ symmetry
 which encodes the exact solutions of the (topological) theory.
 More general ground ring elements can be obtained acting
 on ${\cal G}_n$ with isometries of $H_3^+ \times S^3$, though they will not be in the
$\G_0^+$ cohomology.

\newsubsection{Small phase space deformations}

As we have seen, deformations of the minimal string obtained by
 turning on the first $p-1$ tachyons (the so-called
 ``small phase space'')
 lift to deformations of the {\it state}
 of string theory on $H_3^+ \times S^3$.
 Each point in the small phase space maps
 to a certain 1/2 BPS configuration in $H_3^+ \times S^3$.
These configurations are exact solutions
of tree level string theory, for finite $\alpha'$.
We can write down the corresponding sigma models
in the Wakimoto representation,
\be \label{sigma}
S_{H_3^+ \times S^3} \to S_{H_3^+ \times S^3} +\sum_{n=1}^{p-1} t_n \int d^2 z\,    \, \beta_M \bar \beta_M
e^{2 i \frac{p - 1 - n}{2 \sqrt{p}} x } e^{2 \frac{n-p-1}{2 \sqrt{p}}\phi} \,.
\ee
We see that they are
deformations of the $S^3$, ``gravitationally dressed''
by the warp factor of $H_3^+$.
The underlying integrable structure guarantees that they
are exactly marginal.  They
correspond to states since they preserve the $H_3^+ \times S^3$ asymptotics as $\varphi \to \infty$.

In the limit $p \to \infty$, these sigma models must correspond
 to 1/2 BPS supergravity solutions. Such supergravity
solutions for the NS5/F1 system have been classified
 \cite{Lunin:2001fv,Lunin:2002bj,Lunin:2002iz, Martelli:2004xq, Liu:2004hy},
and the sigma models (\ref{sigma}) provide a generalization to
finite $\alpha'$ for a subclass of them. A natural guess
is that this  class corresponds to (a subsector of) the Coulomb branch
of the near horizon geometry of the NS5/F1 system.
It would be nice to understand the geometric interpretation of (\ref{sigma}) in  more detail.

\newsection{Holography and  symmetric products}

The main interest of string theory on $H_3^+ \times S^3$ is in the
context of the AdS/CFT correspondence. This background arises as
the near horizon geometry of $Q_5$ NS5 branes wrapping $M_4$  and
$Q_1$ parallel fundamental strings. The dual spacetime CFT on the
boundary of $H_3^+$
 is believed to be the low-energy limit of the worldvolume theory
on the NS5-F1 system: a sigma model  with target space a certain
deformation of Sym$^{Q_1 Q_5}(M_4)$.
The string coupling constant $g_s$ is fixed by the relation
\cite{Giveon:1998ns}
 \be
Q_1 \sim {{\rm Vol}_{M_4}\over \alpha'^2 g_s^2 \sqrt{Q_5}} \,.
\ee
For a fixed level $ p \equiv Q_5$ of the worldsheet sigma model,
 string perturbation theory corresponds to $Q_1 \to \infty$.
All the topological  amplitudes that we have considered in this
paper arise  at string tree level, and map to boundary correlators
at leading order in the large $Q_1$ expansion.

While the details of the boundary theory are not very well
understood, the spectrum of spacetime chiral primaries of the
string theory has been matched with the spectrum of chiral  twist
operators of the symmetric product CFT.  It
has been suggested  \cite{Argurio:2000tb} that it is more natural to phrase the
correspondence in terms of Sym$^{Q_1}$(Sym$^{Q_5}(M_4))$, which
has the same chiral spectrum of  Sym$^{{Q_1 Q_5}}(M_4)$ and is
believed to be in the same moduli space. Here one thinks of
Sym$^{Q_5} (M_4)$ as the theory of a single long string. Twist
operators of  Sym$^{Q_5} (M_4)$  in the untwisted sector of
Sym$^{Q_1}$ correspond to ordinary vertex operators for $H_3^+
\times S^3$, with zero amount of spectral flow. The operation of
taking $w$ units of spectral flow in the bulk maps to considering
the $Z_w$ twisted sector of  Sym$^{Q_1}$.

Let us first consider the chiral twist operators  for Sym$^{Q_5} (M_4)$.
  We restrict to universal operators  that do not depend on the structure of $M_4$. They
 were constructed for instance in \cite{Vafa:1994tf,Lunin:2001pw}:
\ba \sigma_n^- : h = {n-1\over 2}  \quad  n = 1, \ldots, Q_5  \eol
\sigma_n^+ : h = {n+1\over 2}  \quad  n = 1, \ldots, Q_5  \ea
We have written the left-movers only, and by left-movers we now mean
left-moving in the boundary CFT. The operator $\sigma_n^\pm$
contains a twist operator of length $n$. The correspondence
between the bulk and boundary theories is
\ba {\cal W}_l & \to & \sigma^-_{2l+1}, \quad  l = 0, \ldots, {p-2
\over 2} \eol {\cal X}_l & \to & \sigma^+_{2l+1}, \quad l = 0,
\ldots, {p-2 \over 2}. \ea
By including the $Z_w$ twisted sector for Sym${^Q_1}$ one  finds  \cite{Argurio:2000tb}
chiral operators with the correct dimensions to be identified with
the spectral flowed bulk vertex operators.

From the bulk viewpoint the spectrum of
dimensions appears to increase indefinitely, while from the boundary viewpoint it is cut-off. The reason for this is that in the bulk we
work in perturbation theory, so we have implicitly taken $g_s
\to 0$. If we want to recover the same results int the boundary
theory we should take $Q_1 \sim 1/g_s^2 \to \infty$, which removes
the cut-off. If we want to consider $Q_1$ finite, then there
must be non-perturbative effects  in the bulk to cut-off the spectrum. This
is the well-known ``stringy exclusion principle'' \cite{Maldacena:1998bw}.

The missing states in the bulk spectrum  correspond
 to the twist operator of length $Q_5$ in
Sym$^{Q_5}(M_4)$ and its images in Sym$^{Q_1}$. In particular,
they are naively present in the boundary theory, and one must
invoke subtle effects \cite{Seiberg:1999xz} to argue them away.

Next we would briefly like to discuss some properties of
amplitudes in the symmetric product theory. The maps from the
boundary of $H_3^+$ to the symmetric product can be described in
terms of a covering surface $\Sigma_g$ of degree $Q_1 Q_5$ of the
boundary, together with a map from this covering surface to $M_4$.
Since our topological string theory was completely defined in
terms of $H_3^+ \times S^3$, the amplitudes we are looking for in
the boundary theory can only depend on the constant modes of
$M_4$. So we can effectively ignore the map $\Sigma_g \to M_4$ and
focus on the structure of the covering $\Sigma_g \to {\bf P}^1$.

Consider an amplitude of various tachyons and
descendants in the bulk. This translates to
computing an amplitude in the boundary theory with certain twist
operator insertions. We further need to
turn on interactions in the symmetric product theory. This is done
by inserting DVV
 \cite{Dijkgraaf:1997vv} twist operators on the boundary and integrating over their positions. Finally to
recover the perturbative answer one should take the limit $Q_1\to
\infty$.

There are many details that need to be worked out in order to
compute such amplitudes. However since we are dealing with a
relatively simple topological theory in the bulk, our guess is
that the amplitudes on the boundary also end up being those of a
well known topological theory, namely Hurwitz theory. That is, we
suspect that up to a numerical factor the amplitudes simply count
the number of covers with the branching we have just described.

The Hurwitz problem has been completely solved. Some interesting
works that might be relevant here are
\cite{Kostov:1997bn,OkounkovToda}. The Hurwitz problem is known to
be related to topological strings \cite{ELSV,Okounkov:2000gx}.

\newsection{Relation with Calabi-Yau spaces and generalizations}

\newsubsection{Calabi-Yau description}

We have seen that $H_3^+ \times S^3$ gives a $\hat{c} = 3$
topological string realisation of the minimal $(p,1)$ string
theories. Another such realisation was recently proposed in
\cite{Aganagic:2003qj}. These authors considered the non-compact
 Calabi-Yau manifold defined by the equation
\be\label{dualCY}
 y + x^p + \dots + uv = 0 \, ,
 \ee
and argued that an appropriately defined  B-model topological
string on this Calabi-Yau is equivalent to the $(p,1)$ minimal
string.\footnote{A certain amount of interpretation is needed, since we
expect the minimal string (which has a linear dilaton)
to describe the leading behaviour near a
Calabi-Yau singularity, whereas
if we regard this equation as an equation
in ${\bf C}^4$ we get no singularities. We will not discuss
this issue here.}
Since this threefold is closely related to the ground ring
of the minimal $(p,1)$ string, one can view this as ``quantization
of the deformations of the ground ring''.  This Calabi-Yau has no
K\"ahler  moduli and therefore the associated A model is trivial.

Our topological theory has the same B model, namely the $(p,1)$
minimal string,
  and in fact it has also a trivial A model. This can be seen easily by
  recalling that $\G^-$ is equivalent to the $B$ antighost of the minimal string,
  which clearly has trivial cohomology. Thus we have two $\hat{c}=3$ $(2,2)$ SCFTs
for which both the A  and B models agree. In such a situation it
is expected that superconformal invariance also forces the D-terms
to agree. So we conclude that the two theories must be identical.
The question then arises if there is a natural explanation for
this equivalence.
We conjecture that $H_3^+ \times S^3$ and the Calabi-Yau
sigma-model are related by T-duality. An example of such a duality
for SCFTs that is very similar in spirit is the equivalence
between the NS5-brane geometry  (which is analogous to  $H_3^+ \times
S^3$) and the ALE sigma model (which is analogous to the Calabi-Yau
threefold). These two CFTs are related by T-duality
\cite{Ooguri:1995wj}.

\newsubsection{$H_3^+ \times S^1$}

It is also  interesting to consider more general backgrounds of
the form $H_3^+ \times {\cal N}$ with NS flux, where ${\cal N}$ is
some coset model. So long as the worldsheet theory has $N=2$
supersymmetry, one can write down a topological string theory and
study its properties. We will now take a look at the simplest of
these examples, namely $H_3^+ \times S^1$. We will take $H_3^+$ at
fractional level $t+2 =p/q+2$ for now and restrict ourselves later
on.

The generators are given as follows:
\ba {\sf T} &=&  - (\del x)^2 - {1\over 2t}\left( 2 k_3 k_3 + k_+
k_- + k_- k_+ \right) \eol
 & &   + \half (\del c^2 b_2  -  c^2 \del b_2) + \half (\del c^1 b_1
 - c^1 \del b_1) +{\kappa \over 2t}\, \del {\sf J}_3^{\rm tot} \eol
{\sf G}^+ &=& c^1 k_-  + c^2(  i\sqrt{t}\, \del x +  k_3 + c^1
b_1)   \eol {\sf G}^- &=& {1\over t}\left( - b_1 k_+ + b_2( i
\sqrt{t}\, \del x -  k_3 - c^1 b_1) + \kappa\, \del b_2 \right)
\eol {\sf J} &=& c^1 b_1+ c^2 b_2 + {2\over t}( k_3 + c^1 b_1 ) +
{\kappa\over t}{\sf J}_3^{\rm tot} \ea
with \be {\rm J}_3^{\rm tot} =  k_3 +  i \sqrt{t}\, \del x + c^1
b_1 = \{ {\sf G}_0^+, b_2 \}. \ee
Here we used the same Wakimoto representation and  OPEs as in
sections 2 and 3, and we used the current $j= i\sqrt{t}\,\del x$
to describe the $S^1$ factor. From the OPEs one finds that the
central charge is given by $\hat{c} = (2t+2)/t$. When we expand
the twisted stress tensor in the Wakimoto representation with
$\kappa = t-1$ we find
\ba {\sf T} + \half \del {\sf J} &=& - \del x^2 + i {t-1 \over
\sqrt{t}} \del^2 x -  \del \varphi^2 - {t+1 \over \sqrt{t}} \del^2
\varphi \eol & & \qquad + c^1 \del b_1 + 2 \del c^1 b_1 + \del c^2
b_2 + \del \beta \gamma  \, ,\ea
which for $t = p/q$ yields exactly the stress tensor for the
$(p,q)$ minimal model coupled to gravity  plus an additional $\{
b_2,c^2,\beta,\gamma \}$ multiplet.

In this realization, $t$ is no longer restricted to be an integer
since there is no flux quantization. On the other hand,  in the
untwisted theory the scalar $x$ does not have a background charge
and we do not have any matter screening operators, but in a
minimal theory we do need a screening operator, so we need to deal
with non-minimal theories. Fortunately there is one special case,
namely $t=1$ (and $\hat{c}=4$), where we do not have any
background charge for $x$ either before or after twisting. In this
case, by arguments very similar to those for $H_3^+ \times S^3$,
we recover the $c=1$ string at self-dual radius. Namely we can
split up the BRST charge as
\be {\sf G}_0^+ = {\sf Q}_1 + {\sf Q}_2 \ee
with
\ba {\sf Q}_1 &=& \oint c^1 k_-  \eol {\sf Q}_2 &=& \oint c^2(
i\sqrt{t} \, \del x +  k_3 + c^1 b_1) \ea
which satisfy $\{ {\sf Q}_i, {\sf Q}_j \} = 0$. The cohomology of
${\sf Q}_2$ is generated by the familiar expressions
\begin{equation}
\begin{array}{rcll}
e^{-2 \Phi/\sqrt{t}} &=& \beta e^{-2 \varphi/\sqrt{t}}  \eol e^{-2
i X/\sqrt{t}} &=& \beta e^{-2 i x/\sqrt{t}}  \eol B &=& b_1 \beta
\eol C &=& c^1 \beta^{-1}
\end{array}
\end{equation}
The similarity transformation is now defined using the operator
\be {\sf L} = -\oint (c^1 k_+^{-1}) \left(b_2(  i \sqrt{t}\, \del
x -  k_3 - c^1 b_1) + \kappa\, \del b_2 \right) \ee
and we find that
\be
 \exp {\rm Ad}({\sf L}){\sf G}^- = {1\over t} (- b_1 k_+ ) =
-{1\over t} B. \ee
and\footnote{ With the same qualifications as in a previous
footnote.}
\be \exp {\rm Ad}({\sf L}){\sf G}^+_0 = {\sf Q}_2 - t\oint C(T_X +
T_\Phi + \half T_{BC}) + \{ {\sf Q}_2, \bullet \}. \ee
Moreover, ${\sf J}$ is invariant under the similarity
transformation. Therefore this topological theory is equivalent to
the $c=1$ string.

The fact that the spectrum of topological string theory on $H_3^+
\times S^1$ coincides with the $c=1$ string is perhaps not too
surprising. Indeed imposing invariance under ${\sf Q}_2$ reduces
$H_3^+ \times S^1$ to the Kazama-Suzuki coset $SL(2)/U(1)$ at
level 3 and ${\sf Q}_1$ coincides with the BRST operator for
topological string theory on the coset, which is of course known
to be equivalent to the $c=1$ string \cite{Mukhi:1993zb}. On the
other hand, the coupling to gravity naively appears to be
different for $H_3^+ \times S^1$ and $SL(2)/U(1)$. The similarity
transformation  shows that they are in fact equivalent.

Since $\hat{c} = 4$, one can ask if there is a Calabi-Yau
four-fold realisation of this conformal field theory. By analogy
with (\ref{dualCY})
there is a natural candidate, namely the Calabi-Yau defined by the
equation
\be y + x_1^2 + x_2^2 + x_3^2 + x_4^2 = 0. \ee

The topological string theories on more general spaces of the form
$H_3^+ \times {\cal N}$ with NS flux should be very similar. The
horizon manifold ${\cal N}$ will get reduced to some matter
theory, and $H_3^+$ should yield the gravitational dressing by the
Liouville field. One may also speculate about a Calabi-Yau dual of
such conformal field theories. A natural guess is that if
$SL(2)/U(1) \times {\cal N}/U(1)$ (modulo GSO projection) is
described by a Calabi-Yau $n$-fold of the form $P(x) = \mu$, then
$H_3^+ \times {\cal N}$ is described by the $n+1$-fold $y^q + P(x)
= 0$, with $q=1$ if the level of $H_3^+$ is an integer.

\newsubsection{Minimal NS5-branes}

Another application of the ideas used in this paper is to the
$N=4$ topological string theory \cite{Berkovits:1994vy} on ALE
spaces, or equivalently \cite{Mukhi:1993zb} near horizon NS5-brane
geometries. This topological string theory computes certain BPS
correlators in six dimensions. The (4,4) superconformal field
theory for the four transverse directions to a stack of $p$
NS5-branes consists of an $SU(2)_{p-2}$ WZW model, a linear
dilaton, and four free fermions. We will now show that this
topological string theory also reduces to the $(p,1)$ bosonic
string.\footnote{ This result was announced at a talk at Stony
Brook in August 2005 \cite{Simons}. The recent papers
\cite{tadashi,niarchos,sahakyan} make a similar claim and check
the correspondence for certain tree level amplitudes. Our approach
proves the correspondence for any scattering amplitude at any
genus.} A relation of this kind was
 anticipated in \cite{Ooguri:1995wj}.
 However the disclaimer mentioned in the introduction
 applies: by the $(p,1)$ bosonic string we
  do not mean the  purely  { topological}   model
  realized {\it e.g.} as the $N=2$ $A_{p-1}$ minimal model
  coupled to topological gravity
(\cite{Dijkgraaf:1991qh} and references therein).

  The main point is that if we throw out the quartet $\{ c^2, b_2, \beta_L,
\gamma_L \}$ of $H_3^+ \times S^3$, we end up with the CHS
geometry. Since this quartet did not give any contribution in
topological correlators except for zero modes, we again end up
with the minimal string. By T-duality, this also describes the
topological string on the $A_{p-1}$ ALE spaces.

The generators of the left-moving
$N=4$ algebra of the CHS background can be taken to be\footnote{For simple comparison with the rest
of the paper, we have labelled the fermions by $c^1,b_1$ and $c^3,b_3$.}
\ba
{\sf T} &=& -  \del\varphi^2 -{1\over \sqrt{p}}\del^2 \varphi
 + {1\over 2p}(2 j_3 j_3 + j_+ j_- + j_- j_+)  \eol
 & & \quad + \half (\del c^3 b_3  -  c^3 \del b_3) + \half (\del c^1 b_1
 - c^1 \del b_1)  \eol
{\sf G}^+ &=& c^1 j_- + c^3(- \sqrt{p}\,\del\varphi +  j_3 + c^1 b_1) - \del c^3  \eol
{\sf G}^- &=& {1\over p}\left(
b_1 j_+ + b_3(\sqrt{p}\, \del\varphi +  j_3 + c^1 b_1) + \del b_3 \right)  \eol
\tilde{{\sf G}}^+ &=&
{1\over p}\left(
c^3 j_+ - c^1( \sqrt{p}\, \del\varphi +  j_3 - c^3 b_3) - \del c^1 \right)  \eol
\tilde{{\sf G}}^- &=& b_3 j_- + b_1(\sqrt{p}\, \del\varphi -  j_3 + c^3 b_3) + \del b_1  \eol
{\sf J} &=& c^3 b_3 + c^1 b_1  \eol
{\sf J}^{++} &=& c^3 c^1  \eol
{\sf J}^{--} &=& b_3 b_1
\ea
We are going to use the same Wakimoto representation as before, i.e. $j_+ = \beta_M$, etc.
Following our previous arguments, we split up the BRST operator ${\sf G}^+_0 = {\sf Q}_1 + {\sf Q}_R$
with
\ba
{\sf Q}_1 &=& \oint c^1 j_-  \eol
{\sf Q}_R &=& \oint c^3 (- \sqrt{p}\,\del\varphi +  j_3 + c^1 b_1) \,.
\ea
We can perform a similarity transformation by
\be
{\sf L} = \oint (c^1 \beta_M^{-1})\left(b_3( \sqrt{p}\, \del\varphi +  j_3 + c^1 b_1) + \del b_3 )\right).
\ee
As usual this yields the expected operators for the $(p,1)$ string:
\ba
\exp({\rm Ad}\, {\sf L}) {\sf G}^+_0 &=& {\sf Q}_R + p {\sf Q}_{\rm Vir} + \{ {\sf Q}_R,\bullet \}  \eol
\exp({\rm Ad}\, {\sf L}) {\sf G}^- &=& {1\over p} B.
\ea
We also find
\be
\exp({\rm Ad}\, {\sf L}) \tilde{{\sf G}}^+ = {1\over p} c^3 j_+ = {1\over p} c^3 \beta_M .
\ee
The cohomology of ${\sf G}^+_0$ yields the physical states of the
$(p,1)$ minimal string, as before. It is easy to check
that these cohomology elements are also in the
kernel of $\tilde{\sf G}^+_0$, as required for observables of the
$N=4$ topological string.\footnote{
It is conceivable that by using
free field representations, we miss some discrete states of the full theory.
Such extra states appear in the analysis of \cite{kps}.
However our free field representation should capture all the asymptotic states,
and hence the minimal string should describe all the scattering amplitudes in
the $N=4$ topological string.} Mimicking our previous
arguments, one finds that the correlation functions also agree
with the minimal string. We will elaborate on aspects of this correspondence
in a future paper \cite{LM}.

The relation to the minimal string leads to a lot of new insight.
For instance the $(2,1)$ minimal string, or equivalently $c=-2$ matter coupled to
gravity is solved by a
supermatrix model \cite{David:1985et,Kostov:1987kt,Klebanov:1990sn}. The authors of
\cite{Klebanov:1990sn} find the following all genus
formula for the free energy as a function of the cosmological constant $\mu$,
\be\label{KWfree}
{\cal F}=-{1\over 6} \mu^3 \log \mu + {1\over 12} \log \,\mu -
\sum_{g=2}^{\infty} {(3g-4)! \over 12^g g!} \mu^{3-3g}\,.
\ee
Notice that this answer is very different from pure topological
gravity, which has ${\cal F}_{\rm top}= -{1\over 6} \mu^3  +
{1\over 12}$. The expression (\ref{KWfree}) can be checked for
$g=0$ and $g=1$ in the continuum worldsheet formulation of the
$(2,1)$ minimal string. The logarithmic terms are seen to arise
from integration over the Liouville zero mode.  In particular,
this implies that there is a non-vanishing 4-point function at
tree level, scaling like $1/\mu$.

An interesting observation, pointed out to us by Marcos Mari\~no,
is that for each genus, the coefficients (\ref{KWfree})
 are precisely the leading singular terms in a $\mu=0$ expansion of a simple polylog expression:
\be
{\cal F}_g (\mu)=-{1\over 12^g g!} {\rm Li}_{4-3g}(e^{-\mu}) =-{1\over 12^g g!} \sum_{d=0}^\infty
d^{3g-4} e^{- d \mu} \,.
\ee
This has the form of a worldsheet instanton expansion, and so we get a
prediction for counting genus $g$ curves of degree $d$ on an $A_1$ ALE space!
Similarly, we expect a matrix model with multiple supermatrices to solve the $(p,1)$ minimal string.
This issue is currently under investigation \cite{LM}.

The genus zero term in the free energy (\ref{KWfree}) can be
understood as follows \cite{little}. Suppose we consider type IIB
on the CHS geometry or equivalently type IIA on the $A_1$ ALE.
Then there are light 1/2 BPS states (W-bosons) coming from wrapped
branes with a mass of order $M^2 \sim \mu$. Now the coefficient of
the $F^4$ term in the effective six-dimensional theory is given by
$\del^4 {\cal F}_0/\del \mu^4$, where $F$ is the field strength of
a $U(1)$ vector multiplet and ${\cal F}_0$ is the genus zero free
energy of the topological string. This coefficient can also be
calculated by a one-loop computation with only charged light 1/2
BPS states running around the loop. Therefore this coefficient
should scale like $M^{-2} \sim \mu^{-1}$ and one deduces that
${\cal F}_0 \sim \mu^3 \log \mu$ \cite{little}.

For higher genus the topological string computes the coefficient
of certain $R^4 F^{4g-4}$ terms in the six-dimensional effective
theory \cite{Berkovits:1994vy}, where $R$ is the curvature tensor.
These terms are chiral and one might hope that the answer is fixed
by a one-loop Schwinger calculation involving the 1/2 BPS
states.\footnote{See
\cite{Antoniadis:1995zn,Vafa:1995ta,Gopakumar:1998ii} for similar
computations for the case of the conifold.} But it turns out that
the Schwinger computation yields an answer for ${\cal F}_1$ that
scales as $\mu^{-1}$ instead of $\log \mu$.\footnote{This
discrepancy has been noticed independently by O. Aharony and D.
Kutasov (unpublished).} On closer inspection this is actually not
a problem: the string worldsheet computation which reduces to the
minimal string is valid for large $\mu$, and the Schwinger
computation is valid when the W-bosons are very light, which is
for small $\mu$. In our setting, there is no non-renormalization
theorem which protects these terms. It is quite likely that there
are additional instanton corrections to the $R^4 F^{4g-4}$ terms.
One might hope that reasoning similar to \cite{BVcon,
GSethi,Green:1997tv,Kiritsis:1997em} fixes these terms completely.

\vspace{1cm}

 {\bf Acknowledgments}

 \medskip

We are grateful to D. Gaiotto for initial collaboration on aspects
of this material.  It is a pleasure to thank N. Berkovits, R. Dijkgraaf, D. Kutasov,
J. Maldacena, M.~Mari\~no, E.~Martinec, L.~Motl, O.~Okounkov,
R.~Pandharipande, M.~Ro\u{c}ek, N. Seiberg, T.~Takayanagi,
C.~Vafa, and especially H. Verlinde for discussions. MW would like
to thank SUNYSB, UvA, Caltech, UCSB, SLAC, MIT, Harvard, CMS
ZheJiang University and CMTP ShangHai for hospitality. This
material is based upon work supported by the National Science
Foundation Grant No. PHY-0243680. Any opinions, findings, and
conclusions or recommendations expressed in this material are
those of the authors and do not necessarily reflect the views of
the National Science Foundation.

\appendix

\renewcommand{\newsection}[1]{
\addtocounter{section}{1} \setcounter{equation}{0}
\setcounter{subsection}{0} \addcontentsline{toc}{section}{\protect
\numberline{\Alph{section}}{{\rm #1}}} \vglue .6cm \pagebreak[3]
\noindent{\bf  \thesection. #1}\nopagebreak[4]\par\vskip .3cm}
\renewcommand{\newsubsection}[1]{
\addtocounter{subsection}{1}
\addcontentsline{toc}{subsection}{\protect
\numberline{\Alph{section}.\arabic{subsection}}{#1}} \vglue .4cm
\pagebreak[3] \noindent{\it \thesubsection.
#1}\nopagebreak[4]\par\vskip .3cm}

\newsection{Appendix: Double Complexes}

In this Appendix we justify equation (\ref{total}). We need  a simple lemma
about double-complexes \cite{Bott&Tu}:

Consider the double complex generated by
the differentials $\Q_H$ and $ \Q_V$,
with $\Q_H^2 = \Q_V^2 = \{  \Q_H, \Q_V\} = 0$.
The differentials act on the vector space $$K = \bigoplus_{h, v \in \mathbb{Z}} K^{h, v}\, ,$$
double-graded by the ghost numbers $h$ and $v$:
\ba
Q_H: &\;& K^{h, v} \to K^{h+1, v} \nonumber\\
Q_V: &\; & K^{h, v} \to K^{h, v+1}\,. \nonumber
\ea
The total differential $\Q = \Q_H + \Q_V$
acts on the complex graded by the total ghost
number, $K = \bigoplus_t K^t \equiv \bigoplus_{h+v = t} K^{h, v}$.
With these definitions, we have:
  If the horizontal sequence is exact except
 at most two consecutive gradings, say  $h = 0\, , 1$, {\it i.e.} $ H_{\Q_H}( K^{h, v} ) = 0$
for $h \neq 0 \, , 1$;
then $H_{\Q_H +\Q_V}(  K )  \cong H_{\Q_V} ( H_{\Q_H}( K))$.
This statement follows immediately by considering
the associated spectral sequence, which converges after one step \cite{Bott&Tu}.

The claim (\ref{total})  is easily proved by repeated application of the lemma. The first
equality follows if we take  $\Q_V= {\sf G}^+_0$ and
$\Q_H = \Q_F^M + \Q_F^L$. The exactness of the horizontal sequence
except at zero grading is a famous fact of the Felder construction,
as we reviewed in the text. The second equality follows
if we take $\Q_H = \Q_2$, $\Q_V = \Q_1 + \Q_F^M + \Q_F^L$.
The hypothesis holds since for a given choice of $\beta_L \gamma_L$ picture, the cohomology at $\Q_2$  is non-trivial only for one value of the $q_2$ grading. Finally the last equality follows
with $\Q_H = \Q_3$, $\Q_V = \Q_1 + \Q_F^M + \Q_F^L$.  Here the lemma
can be applied because as explained in section 3.5
the $\Q_3$ cohomology is non-trivial only for $q_3 =0$ and $q_3 =1$.

\newsection{Appendix: Facts about minimal strings}

In this Appendix we review some standard facts about the $(p,q)$
models coupled to gravity, emphasizing their free field description.
The main goal is to point out the special features of the $(p,1)$ models.
Useful reviews with a different perspective include \cite{Dijkgraaf:1990qw,Dijkgraaf:1991qh,Lerche:1993xf}.

\newsubsection{The $(p,q)$ minimal models and their Felder description}

The Virasoro minimal models are labeled by a pair or relatively
prime integers $(p,q)$, $p, q >1$. They have central charge
\be c_{p,q} = 1 - 6 Q_{p,q}^2   \, , \quad  Q_{p,q} \equiv \sqrt{\frac{p}{q}} -\sqrt{ \frac{q}{p}}\,.
\ee
There are $(p-1)(q-1)/2$ primary fields  labeled by two integers $(r,s)$ satisfying $1 \leq r \leq p-1$, $1\leq s \leq q-1$,
with the identification $(r,s) \sim (p-r, q-s)$.
Their conformal dimensions are
\be
\Delta_{r,s} = {(qr - ps)^2 - (p-q)^2 \over 4pq} \, .
 \ee
The Coulomb gas description of these models is in terms
of the theory of free boson $X$ with background charge, supplemented
by appropriate screening operators. The stress tensor reads ($\alpha' = 1)$
\be{\sf T}_X   = - \del X \del X + {i } Q_{p,q} \del^2 X   \,.
\ee
Vertex operators are given by
\be V_\alpha = e^{ 2 i \alpha X} \ee
and have conformal dimension
\be \Delta_\alpha =  \alpha(\alpha-Q_{p,q}).
\ee
It is useful to introduce the lattice of momenta
\be \alpha_{m,n} = \half(1-m)\alpha_- + \half(1-n) \alpha_+ \quad
{\rm with } \quad \alpha_+= \sqrt{{p\over q}}, \quad \alpha_- =
-\sqrt{{q\over p}}  \, .\ee
The conformal dimensions of the corresponding operators $V_{m,n} = e^{2 i \alpha_{m,n} X}$
can be expressed as
\be \Delta_{\alpha_{m,n}} = {1\over 4}( m\alpha_- + n\alpha_+)^2 -
{1\over 4} Q_{p,q}^2   \, .\ee
Note the invariance of $\Delta_{m, n}$  under $(m, n) \to (-m, -n)$ and
$(m , n) \to (m +p, n +q)$.
The spaces $\F_{m,n}$ are defined as the Fock spaces
built on the $(m,n)$ vacua,
\be
\F_{m,n} \equiv {\rm Span} \{ a_{-n_1}  \cdots a_{-n_k}   V_{m,n} (0)  | 0 \rangle \} \, ,
\ee
where  the $a_n$'s denote the usual oscillators, $i \partial X (z)= \sum_n a_n z^{-n +1} $,
and $| 0 \rangle$ is the $SL(2)$ invariant vacuum.
Finally we introduce the conformally invariant screening charges
\be \label{QFM}
\Q_\pm  = \oint e^{2 i\alpha_\pm X}  \,. \ee
With  all the ingredients  in place we can now
present the free field resolution of the irreducible Virasoro module $\L_{r,s}$.
The sequence
\be \label{seq}
  \dots   \stackrel{\Q_-^r}{\longrightarrow}  \F_{2p-r, s}  \stackrel{\Q_-^{p-r}}{\longrightarrow}
  \F_{r, s} \stackrel{\Q_-^r}{\longrightarrow}
  \F_{-r, s}   \stackrel{\Q_-^{p-r}}{\longrightarrow}    \dots
\ee
is a complex, {\it i.e.}  $\Q_F^2 = 0$, where the Felder
BRST charge $\Q_F$ is defined as  ${\Q_-^r}$ or ${\Q_-^{p-r}}$
according to which space it acts on.
Felder proved that the sequence is exact except
on the middle Fock space $\F_{r,s}$, where the
 cohomology $H_{\Q_F}(  \F_{r,s})$ is isomorphic to the irreducible
 Virasoro module ${\cal L}_{r,s}$ \cite{Felder:1988zp}. This construction has the following rationale:
The reducible representation $\F_{r,s}$ contains two primitive
 submodules,  one built on the null at level
$ rs$ and the other built on the null at level $(p-r)(q-s)$.
Restricting to ${\rm Ker}_{ \Q_F}(\F_{r,s})$ factors out  the null at level $rs$, while modding out by ${\rm Im}_{ \Q_F}(\F_{r,s})$
factors out the null at level $(p-r)(q-s)$.

An equivalent resolution is obtained by considering the dual Fock space, $(r,s) \to (p-r, q-s)$.

\newsubsection{Liouville }

The Liouville field $\Phi$ has stress tensor
\be {\sf T_\Phi} = -\del \Phi \del \Phi -
\tilde{Q}_{p,q} \del^2\Phi \ee
and central charge
\be c_\Phi = 1 + 6 \,  \tilde{Q}_{p,q}^2\, , \quad
\tilde{Q} = \sqrt{{p\over q}}  +  \sqrt{{q\over
p}} \,. \ee
Vertex operators of the form
\be W_\beta = e^{2 \beta \Phi} \ee
have conformal dimensions
\be \tilde \Delta_\beta =  \beta(-\beta -\tilde{Q}_{p,q} ) \, . \ee
We are adopting  almost the same conventions as in the recent Liouville literature
({\it e.g.} \cite{Teschner:2001rv, Kostov:2003uh, Seiberg:2003nm}
with the  exception that for us the  weak coupling region is at $\Phi = \infty$.
The Seiberg bound is
\be
\beta \geq -  \frac{\tilde Q_{p,q}}{2} \,.
\ee
For special values of $\beta$ one encounters degenerate
representations:
\be \beta_{m,n} =- \half(1-m)\beta_- - \half(1-n) \beta_+ \quad
{\rm with } \quad \beta_+ = \sqrt{{p\over q}}, \quad \beta_- =
\sqrt{{q\over p}} \ee
with
\be \tilde \Delta_{\beta_{mn}} =-{1\over 4}(m\beta_- + n\beta_+)^2 +
{1\over 4} \tilde{Q}^2 \, . \ee
Next we would like to introduce vertex operators  $e^{2 \gamma_{mn}\Phi}$ with
\be
\gamma_{m,n} =  -\half(1-m)\beta_- - \half(1+n)\beta_+  \, .
\ee
and dimension
\be
 \tilde \Delta_{\gamma_{m,n}} = \tilde \Delta_{\beta_{m,n}} + mn =
-\Delta_{\alpha_{m,n}} + 1
\ee
Note the invariances  of $ \tilde \Delta_{\gamma_{m,n}}$
under $(m,n) \to (-m, -n)$, $(m,n) \to (m+p, n+q)$.
Finally define
\be
\gamma^+_{mn} = {\rm max} (\gamma_{m,n}, \gamma_{-m,-n})\, , \qquad \gamma^-_{mn} = {\rm min} (\gamma_{m,n}, \gamma_{-m,-n})\, .
\ee
The Liouville momentum $\gamma_{r ,s}^+$, which satisfies the Seiberg bound,
 is used to dress the $(r,s)$ matter
primary to obtain a tachyon vertex operator,
 \be
 T_{r,s} = e^{2 i \alpha_{r,s} X}e^{2 \gamma^+_{r,s}\Phi}\, C \bar C \,.
 \ee

 \newsubsection{Cohomology of the $(p,q)$ models}

 Physical states of minimal string theory correspond to cohomology
 classes of the BRST operator $\Q_{\rm Vir}$.
 Let us briefly review the situation for the ordinary $(p,q)$ models with $p, q >1$.
 Denote by $gh$ the $BC$ ghost-number, in conventions where the $SL(2)$ vacuum $|0\rangle$
 has $gh =0$, and $gh(C) = +1$, $gh(B) = -1$. Consider first
 the chiral  (left-moving or right-moving)
  cohomology relative to $B_0$, that is, evaluated
  in the complex of states annihilated by $B_0$.
   The states that obey the Seiberg bound are:

$gh=1$: {\it Tachyon states}.
 These are the $(p-1)(q-1)/2$ states of the form
\be
 T_{r,s} = e^{2 i \alpha_{r,s} X(0)}e^{2 \gamma^+_{r,s}\Phi(0)} \, C\,.
 \ee

$gh \leq 0$: {\it Lian-Zuckerman states}.
 In correspondence
 to each matter null, there is a LZ state of zero or negative ghost number,
with Liouville dressing equal to that of the corresponding null.
In the free field realization, LZ states can be found
by  a descent procedure, starting with a state $|T_{r + (k+1) p ,\,s} \rangle$,
$1 \leq r \leq p-1$, $1 \leq s \leq q-1$,  $k \geq 0$,
that is, a ``tachyon''  built with a matter primary outside the minimal Kac table.
The descent is \cite{Lian:1991gk,Bouwknegt:1991yg,Imbimbo:1991ia,Govindarajan:1992kv}
\ba \label{descent}
\Q_{\rm Vir} |LZ \rangle^{(-k)} &  = &\Q_F |I_1 \rangle^{(-k+1)} \\
\cdots \cdots \nonumber \\
\Q_{\rm Vir}  |I_{k} \rangle^{(0)} &  = &(-1)^{k} \Q_F |T_{m,n} \rangle^{(1)} \, ,\nonumber
\ea
where the superscripts indicate the ghost number.
Notice that here we are {\it not} imposing the restriction
to half of the Kac table, in other terms we have $(p-1)(q-1)$ states
for each $gh \leq 0$. This doubling is due to the fact
that over each matter primary there are {\it two} primitive null vectors,
each of which generates a LZ state upon descent.
(Applying literally the above descent procedure, half of the LZ states will
 land in the dual matter representation - they can be dualized
back to the usual representation if desired.)

The LZ states with $gh = 0$ form the  {\it ground ring}, generated by \cite{Kutasov:1991qx}
\ba \label{xy}
\hat x  & = & (BC +  \sqrt{p/q}(\del \Phi - i \del X)) e^{
\sqrt{\frac{q}{p}} (\Phi + i X)   }
 \nonumber \\ \hat y &  = & [BC + \sqrt{q/p}(\del \Phi + i \del X)]
e^{\sqrt{\frac{p}{q}}(\Phi - i X)}
\, ,
\ea
subject to the
relations (taking  zero cosmological constant for simplicity)
\be
\hat x^{p-1} =0 \,   , \quad \hat y^{q-1} = 0 \,.
\ee
Cohomology  spaces at given ghost number form representations
of the ground ring. The representation is faithful
for the states with $gh \leq 0$, but not for the tachyons.
When acting on the tachyon module, the ground
ring generators obey the additional relation $\hat x^{p-2} =\hat  y^{q-2}$.

The states listed so far represent only half of the relative chiral
cohomology. Because of the pairing induced by the bpz
inner product, $\langle \Psi | C_0 | \Psi' \rangle$, to each of
the above states corresponds a dual state with
ghost number $2 -gh$ and dual matter and Liouville momenta.
The dual states violate the Seiberg bound.

In building closed string states, we need to combine
left and right movers. We are instructed to work in the semi-relative
cohomology $H_S$,  the cohomology of $\Q_{\rm Vir} + \bar \Q_{\rm Vir}$
relative to $B_0 - \bar B_0$. Consider first the
closed string cohomology $H_R$, relative to {\it both} $B_0$ and $\bar B_0$.
These are the traditional closed string states obtained by gluing
left- and right-movers belonging to the relative chiral cohomology
discussed above. Then one can show \cite{Witten:1992yj}
that $H_S^{(n)} \cong H_R^{(n)} \oplus
H_R^{(n-1)}$. Each representative $\psi^{(n)}$ in the relative
cohomology  $H_R^{(n)}$ of ghost number $n$ gives rise
to two representatives in semi-relative cohomology: one
in $H_S^{(n)}$ immediately given by  $\psi^{(n)}$,
and one in $H_S^{(n+1)}$ of the form $(C_0 + \bar C_0) \psi^{(n)} + \dots$.

The discussion of semi-relative cohomology is more than
a technical nuisance, in fact it important
for the construction of symmetry currents.
For the $c=1$ string, one can obtain symmetry
currents by descent from physical states of (left, right) ghost number $(1, 0)$ made
by gluing a left-moving
tachyon with a right-moving ground ring state.
For the $(p,q)$ models  such physical states are not allowed since the left and right Liouville
momenta do not match.  Fortunately there are states of total ghost number one
  in $H_S^{(1)}$ of the form $(C_0 + \bar C_0) \psi^{(0)}   + \dots$
 where $\psi^{(0)}$ is a ground ring state. As explained in \cite{Witten:1992yj},
 these more general states of ghost number one still lead
 to symmetry currents.

\newsubsection{The $(p,1)$ models}

We are mainly interested in the $(p,1)$ models, which are
outside the range of definition of the ``minimal'' Virasoro series, since
the fundamental domain of the Kac table is empty. Nevertheless
it is possible to construct consistent CFTs with central charge $c_{p,1}$
\cite{Kausch:1990vg,Kausch:1995py,Flohr:1995ea}
These models share with the minimal series the
property of containing only degenerate Virasoro representations - but infinitely many of them, of course.
They are rational with respect to an extended W-algebra \cite{Kausch:1990vg,Flohr:1995ea}.
Modular invariant partition functions have also been constructed \cite{Flohr:1995ea}.

 In the $(p,1)$ model the structure of the degenerate Virasoro representations is somewhat
 different.  Without loss of generality we can label the degenerate Virasoro representations
of a theory with central charge $c_{p,1}$
by  pairs of integers $(r,s)$ with restrictions $ 1\leq r \leq p$ and $s \geq  1$.
Consider again the complex (\ref{seq}), for a theory of central charge $c_{p,1}$,
 and with the understanding that now $(r,s)$ obey the new restrictions.
 One can prove that the sequence is exact, and that the irreducible Virasoro module
 $\L_{r,s}$ is  isomorphic  to ${\rm Ker}_{ \Q_F}(\F_{r,s})$. This is because the reducible representation
 $\F_{r,s}$ has only {\it one} primitive submodule, built on the null at level
 $rs$; the other putative null is absent since it would appear at level $(p-r)(1-s) \leq 0$.

We are going to define the (chiral sector of the) $(p,1)$ model
as containing each irreducible representation $\L_{r, s}$
with $1 \leq r \leq p-1$, $s \geq 1$ precisely once. Notice
that representations with $r = 0$ mod $p$ are excluded. This
is the natural definition that makes contact with the $N=2$ topological
minimal models and the $p$-KdV integrable hierarchy. It is also the
definition that emerges in our reduction from $H_3^+ \times S^3$.

The existence of infinitely many matter primaries has the
effect of changing the structure of the cohomology.
In essence, the putative LZ states with $gh < 0$ are
replaced by states with $gh = 0$ and $gh=1$.
The states in the (chiral, relative) cohomology obeying the Seiberg bound are:

$gh=1$: {\it Tachyon states},
\be
T_{r,s}  = e^{2 i \alpha_{r,s} X} e^{2 \gamma^+_{r,s}  \Phi} C \, , \quad  1 \leq r \leq  p-1 \,  , \quad s  \geq 1\, ,
\ee
one for each matter primary.
For convenience we may relabel them using a single index $n  \equiv p s - r$,
\be
T_n  = e^{2 i   \frac{p-1-n}{2 \sqrt{p}}  X} e^{2  \frac{n-p-1}{2 \sqrt{p} }  \Phi} C \, , \quad n \geq 1\, , \;  n \neq 0 \; {\rm  mod} \;p  \,.
\ee

$gh = 0$: {\it Ground ring states},
\be
G_{r,s} = \hat x^{r-1} \hat  y^{s-1} \, ,\quad  1 \leq r \leq  p-1 \, , \quad   s \geq 1 \, ,
\ee
where $\hat x$ and $\hat y$ are given
by the same expression (\ref{xy}) with $q=1$.
The $\hat x$ generator obeys the usual relation $\hat x^{p-1} = 0$
but there is no restrictions  on the power of $\hat y$. Since each matter
primary has a single primitive null, there is precisely one
ground ring state for each tachyon. States
with $gh < 0$ are absent:  the descent
procedure of the previous subsection always terminates after one step,
on a ground ring state.

As in the $(p, q)$ case, each of the above states has a dual
(of ghost number $2 - gh$) violating the Seiberg bound.

In the $(p,1)$ models the tachyons and the ground ring states
form two isomorphic modules of the ground ring.
 In the formulation of the $(p,1)$ model as twisted
$N=2$ minimal matter coupled to topological gravity,
these two modules should be viewed as
equivalent copies of the (gravitationally extended) chiral ring. In that language,
the states with $s=1$ span the so-called ``small phase space'',  while the states with $s>1$ are interpreted as
 gravitational descendants.

The discussion of the closed string semi-relative cohomology
exactly parallels that for general $(p,q)$.

Finally, we would like to express the view that the most natural way to
treat the usual $(p,q)$ minimal strings with $q >1$
should parallel the above analysis of the $(p,1)$ models -
one should define the matter theory so that it contains an infinite number of  primaries.
 Indeed by coupling to gravity, the fusion
rules of the $(p,q)$ matter minimal model get erased - the zeroes
of the matter correlators are offset by infinities in the Liouville
sector \cite{Dotsenko:1991id} - and it seems necessary to go outside the minimal Kac table.
This treatment of the $(p,q)$ models is also natural from the viewpoint of
obtaining them by gravitational RG flow starting from the $c=1$ theory.

\end{document}